\documentclass[useAMS,usenatbib]{mn2e}

\usepackage{float}
\usepackage{morefloats}
\usepackage{graphicx}
\usepackage{url}
\usepackage[english]{babel}
\usepackage{placeins}

\title[Saturn's Irregular Satellites - Dynamics]{A new perspective on the irregular satellites of Saturn - I\\Dynamical and collisional history}
\author[Turrini et al.]{D. Turrini$^{1,2}$\thanks{E-mail: diego.turrini@ifsi-roma.inaf.it}, F. Marzari$^{2}$, H. Beust$^{3}$\\
$^{1}$Center of Studies and Activities for Space ``G. Colombo'', University of Padova, Via Venezia 15, 35131, Padova, Italy\\
$^{2}$Physics Department, University of Padova, Via Marzolo 8, 35131, Padova, Italy\\
$^{3}$Laboratoire d'Astrophysique de Grenoble, UMR 5571 CNRS, Universit\'e J. Fourier, BP 53, 38041 Grenoble Cedex 9, France\\}
\begin{document}

\date{Accepted XXX. Received XXX; in original form XXX}

\pagerange{\pageref{firstpage}--\pageref{lastpage}} \pubyear{2007}

\maketitle

\label{firstpage}

\begin{abstract}
The dynamical features of the irregular satellites of the giant planets argue against an in-situ formation and are strongly suggestive of a capture origin. Since the last detailed investigations of their dynamics, the total number of satellites have doubled, increasing from $50$ to $109$, and almost tripled in the case of Saturn system. We have performed a new dynamical exploration of Saturn system to test whether the larger sample of bodies could improve our understanding of which dynamical features are primordial and which are the outcome of the secular evolution of the system. We have performed detailed N--Body simulations using the best orbital data available and analysed the frequencies of motion to search for resonances and other possible perturbing effects. We took advantage of the Hierarchical Jacobian Symplectic algorithm to include in the dynamical model of the system also the gravitational effects of the two outermost massive satellites, Titan and Iapetus. Our results suggest that Saturn's irregular satellites have been significantly altered and shaped by the gravitational perturbations of Jupiter, Titan, Iapetus and the Sun and by the collisional sweeping effect of Phoebe. In particular, the effects on the dynamical evolution of the system of the two massive satellites appear to be non-negligible. Jupiter perturbs the satellites through its direct gravitational pull and, indirectly, via the effects of the Great Inequality, i.e. its almost resonance with Saturn. Finally, by using the Hierarchical Clustering Method we found hints to the existence of collisional families and compared them with the available observational data.
\end{abstract}

\begin{keywords}
planets and satellites: formation, Saturn, irregular satellites - methods: numerical, N--Body simulations - celestial mechanics
\end{keywords}

\section{Introduction}

\begin{table*}
\begin{center}
\caption{Mean orbital elements for the $35$ irregular satellites of Saturn computed with Model $1$. For each orbital element, we report the minimum and maximum values attained during the simulations.}
\label{proper-elements1}
\begin{tabular}{lccccccccc}
\hline
\\
& Mean & Max & Min & Mean & Max & Min & Mean & Max & Min \\
\textbf{Satellite} & $a$ & $a$ & $a$ & $e$ & $e$ & $e$ & $i$ & $i$ & $i$ \\
& ($10^6$ km) & ($10^6$ km) & ($10^6$ km) & & & & (degree) &  (degree) &  (degree) \\
\\
\hline
\\
Ijiraq    & 11.34999 & 11.41432 & 11.29464 & 0.27844 & 0.58154 & 0.06807 & 48.539258 & 54.56960 & 38.76520\\
Kiviuq    & 11.36345 & 11.42928 & 11.30960 & 0.22697 & 0.59759 & 0.00007 & 48.926008 & 55.12770 & 38.22650\\
Phoebe    & 12.93274 & 13.01501 & 12.86542 & 0.16320 & 0.18957 & 0.13817 & 175.04998 & 177.9515 & 172.4591\\
Paaliaq   & 15.00018 & 15.21410 & 14.79523 & 0.34480 & 0.65704 & 0.10685 & 49.265570 & 56.93310 & 38.26580\\
Skathi    & 15.57164 & 15.76762 & 15.39362 & 0.27497 & 0.38604 & 0.17590 & 151.94091 & 157.0688 & 146.6004\\
Albiorix  & 16.32861 & 16.69512 & 16.03689 & 0.47653 & 0.64654 & 0.32214 & 37.163084 & 45.95380 & 28.12050\\
S/2004 S11& 16.96589 & 17.38327 & 16.63528 & 0.47777 & 0.65718 & 0.31552 & 38.055491 & 47.02570 & 28.69770\\
S/2006 S8 & 17.53287 & 17.90687 & 17.24863 & 0.49102 & 0.60971 & 0.37623 & 158.94755 & 165.8655 & 151.1970\\
Erriapo   & 17.53885 & 18.02654 & 17.14392 & 0.47078 & 0.64729 & 0.31049 & 37.244669 & 46.09640 & 28.22680\\
Siarnaq   & 17.82459 & 18.35566 & 17.41319 & 0.30400 & 0.62649 & 0.06484 & 47.848986 & 55.71640 & 38.58300\\
S/2004 S13& 18.05048 & 18.43046 & 17.75727 & 0.25649 & 0.33598 & 0.18553 & 168.73717 & 172.2910 & 164.9780\\
S/2006 S4 & 18.05198 & 18.46038 & 17.74231 & 0.32108 & 0.39896 & 0.24616 & 174.23306 & 177.5732 & 170.9701\\
Tarvos    & 18.18362 & 18.80445 & 17.65255 & 0.52954 & 0.72046 & 0.35404 & 37.828226 & 48.25900 & 27.21730\\
S/2004 S19& 18.40203 & 18.77453 & 18.10134 & 0.32516 & 0.48143 & 0.18885 & 149.89285 & 156.3719 & 143.0364\\
Mundilfari& 18.55462 & 18.95405 & 18.22102 & 0.21832 & 0.30069 & 0.14959 & 167.02486 & 170.6053 & 163.2256\\
S/2006 S6 & 18.63391 & 19.01389 & 18.32574 & 0.21771 & 0.29609 & 0.14597 & 162.73732 & 166.6379 & 158.6233\\
S/2006 S1 & 18.82839 & 19.16349 & 18.53518 & 0.12479 & 0.19673 & 0.06475 & 156.27335 & 160.0706 & 152.3567\\
Narvi     & 19.31458 & 19.83668 & 18.90917 & 0.40869 & 0.67753 & 0.18973 & 142.37869 & 153.1620 & 132.3253\\
S/2004 S17& 19.38938 & 19.80676 & 19.04381 & 0.17873 & 0.24785 & 0.11784 & 167.71154 & 171.1124 & 164.2361\\
Suttungr  & 19.39985 & 19.77684 & 19.07373 & 0.11514 & 0.16560 & 0.07043 & 175.87422 & 178.8399 & 173.3134\\
S/2004 S15& 19.61378 & 20.01620 & 19.25325 & 0.14749 & 0.22541 & 0.08185 & 158.77860 & 162.5818 & 154.7763\\
S/2004 S10& 19.68409 & 20.25555 & 19.25325 & 0.27380 & 0.36781 & 0.18218 & 165.90811 & 169.9454 & 161.5303\\
S/2004 S12& 19.70653 & 20.28547 & 19.26821 & 0.33811 & 0.45731 & 0.23433 & 164.64473 & 169.1534 & 159.5224\\
S/2004 S09& 20.20170 & 20.70435 & 19.79180 & 0.24366 & 0.36087 & 0.14068 & 156.17912 & 160.9251 & 150.9430\\
Thrymr    & 20.39019 & 21.07834 & 19.82172 & 0.45790 & 0.57133 & 0.34482 & 175.17339 & 179.0080 & 171.3722\\
S/2004 S14& 20.51136 & 21.13818 & 20.06107 & 0.35780 & 0.46894 & 0.25126 & 165.04688 & 169.6972 & 159.7623\\
S/2004 S18& 20.58018 & 21.93105 & 19.73196 & 0.52264 & 0.80486 & 0.27015 & 138.44452 & 153.5911 & 125.2909\\
S/2004 S07& 20.93622 & 21.70665 & 20.37523 & 0.52895 & 0.66062 & 0.39225 & 163.97039 & 170.4156 & 156.1659\\
S/2006 S3 & 20.97661 & 21.72161 & 20.43507 & 0.46106 & 0.61835 & 0.31309 & 156.61094 & 164.0096 & 148.0057\\
S/2006 S2 & 21.86074 & 22.82864 & 21.15314 & 0.51146 & 0.70784 & 0.31893 & 153.00277 & 162.3733 & 142.1259\\
S/2006 S5 & 22.63865 & 23.60654 & 21.90113 & 0.27610 & 0.46116 & 0.11888 & 167.66132 & 171.7822 & 162.8017\\
S/2004 S16& 22.89895 & 23.65142 & 22.26016 & 0.15618 & 0.24896 & 0.07682 & 164.71501 & 168.4223 & 160.7656\\
S/2006 S7 & 22.90643 & 23.93566 & 22.18536 & 0.44673 & 0.58911 & 0.30290 & 168.38375 & 173.6354 & 161.7933\\
Ymir      & 23.02910 & 24.02542 & 22.30504 & 0.33742 & 0.45724 & 0.22175 & 173.05857 & 176.9269 & 169.0612\\
S/2004 S08& 24.21092 & 25.23716 & 23.39711 & 0.21886 & 0.32816 & 0.11758 & 170.16930 & 173.7150 & 166.4152\\
\\
\hline 
\end{tabular}
\end{center}
\end{table*}
\begin{table*}
\begin{center}
\caption{Mean orbital elements for the $35$ irregular satellites of Saturn computed with Model $2$. For each orbital element, we report the minimum and maximum values attained during the simulations.}
\label{proper-elements2}
\begin{tabular}{lccccccccc}
\hline
\\
& Mean & Max & Min & Mean & Max & Min & Mean & Max & Min \\
\textbf{Satellite} & $a$ & $a$ & $a$ & $e$ & $e$ & $e$ & $i$ & $i$ & $i$ \\
& ($10^6$ km) & ($10^6$ km) & ($10^6$ km) & & & & (degree) &  (degree) &  (degree) \\
\\
\hline 
\\ 
Kiviuq    & 11.34550 & 11.42928 & 11.27968 & 0.26136 & 0.58566 & 0.00007 & 48.144251 & 54.30030 & 38.17760\\
Ijiraq    & 11.35149 & 11.42928 & 11.27968 & 0.30076 & 0.56849 & 0.10084 & 48.085587 & 54.21280 & 39.00740\\
Phoebe    & 12.94620 & 13.02997 & 12.88038 & 0.16190 & 0.18862 & 0.13691 & 175.04566 & 177.9671 & 172.4479\\
Paaliaq   & 14.94632 & 15.19914 & 14.70547 & 0.34683 & 0.65790 & 0.10713 & 49.179928 & 56.89150 & 38.03960\\
Skathi    & 15.59707 & 15.78258 & 15.43850 & 0.27499 & 0.38629 & 0.17705 & 151.77871 & 157.0443 & 146.1022\\
Albiorix  & 16.25381 & 16.68016 & 15.87233 & 0.47472 & 0.64581 & 0.31928 & 37.148303 & 45.80770 & 28.10180\\
S/2004 S11& 16.90157 & 17.41319 & 16.44081 & 0.47628 & 0.66072 & 0.30682 & 38.199374 & 47.39460 & 28.68510\\
Erriapo   & 17.51193 & 18.07142 & 17.08408 & 0.46727 & 0.64520 & 0.30602 & 37.237629 & 45.98280 & 28.28960\\
S/2006 S8 & 17.59271 & 18.02654 & 17.26359 & 0.48969 & 0.61015 & 0.37243 & 158.90749 & 165.8853 & 151.1704\\
Siarnaq   & 17.80065 & 18.31078 & 17.38327 & 0.30301 & 0.61990 & 0.06262 & 47.806963 & 55.69950 & 38.75980\\
S/2006 S4 & 18.00261 & 18.38558 & 17.72735 & 0.32398 & 0.39749 & 0.25221 & 174.25662 & 177.5717 & 171.0669\\
Tarvos    & 18.03702 & 18.86429 & 17.39823 & 0.52401 & 0.71599 & 0.34681 & 37.896709 & 48.34880 & 27.33550\\
S/2004 S13& 18.05945 & 18.43046 & 17.74231 & 0.25507 & 0.33215 & 0.18584 & 168.79586 & 172.3424 & 165.1410\\
S/2004 S19& 18.39156 & 18.75957 & 18.08638 & 0.32258 & 0.48127 & 0.18605 & 149.83505 & 156.3264 & 143.1073\\
Mundilfari& 18.60848 & 18.96901 & 18.31078 & 0.21235 & 0.28177 & 0.14879 & 167.07964 & 170.6035 & 163.4678\\
S/2006 S6 & 18.65485 & 19.01389 & 18.35566 & 0.21690 & 0.29538 & 0.14741 & 162.70489 & 166.6243 & 158.5058\\
S/2006 S1 & 18.83736 & 19.16349 & 18.53518 & 0.12457 & 0.19678 & 0.06425 & 156.26551 & 160.0878 & 152.3155\\
Narvi     & 19.25175 & 19.73196 & 18.86429 & 0.42033 & 0.71140 & 0.19685 & 141.70617 & 152.9912 & 130.2114\\
Suttungr  & 19.36694 & 19.74692 & 19.02885 & 0.11563 & 0.16592 & 0.07091 & 175.87927 & 178.8359 & 173.3126\\
S/2004 S17& 19.37741 & 19.79180 & 19.01389 & 0.17944 & 0.24828 & 0.11906 & 167.73842 & 171.1409 & 164.2874\\
S/2004 S15& 19.60929 & 20.01620 & 19.25325 & 0.15033 & 0.22944 & 0.08377 & 158.75916 & 162.5991 & 154.7317\\
S/2004 S12& 19.69905 & 20.27051 & 19.25325 & 0.33479 & 0.44865 & 0.22925 & 164.65164 & 169.1637 & 159.6312\\
S/2004 S10& 19.78731 & 20.28547 & 19.40284 & 0.26522 & 0.35715 & 0.18270 & 165.94975 & 169.9036 & 161.7321\\
S/2004 S09& 20.19122 & 20.70435 & 19.77684 & 0.24617 & 0.36485 & 0.14408 & 156.23885 & 161.0478 & 150.9973\\
S/2004 S18& 20.20170 & 20.86890 & 19.70204 & 0.50176 & 0.78185 & 0.25099 & 138.17034 & 152.6666 & 125.7233\\
Thrymr    & 20.33783 & 21.09330 & 19.80676 & 0.47665 & 0.59183 & 0.35765 & 175.13448 & 179.0279 & 171.3395\\
S/2004 S14& 20.50837 & 21.12322 & 20.06107 & 0.35662 & 0.46776 & 0.25058 & 165.05483 & 169.7195 & 159.8302\\
S/2004 S07& 20.93473 & 21.70665 & 20.37523 & 0.52853 & 0.66139 & 0.39330 & 163.94112 & 170.4116 & 156.0144\\
S/2006 S3 & 21.08432 & 21.93105 & 20.47995 & 0.46378 & 0.63240 & 0.30706 & 156.66042 & 164.1352 & 147.7268\\
S/2006 S2 & 21.87121 & 22.82864 & 21.18306 & 0.49223 & 0.68729 & 0.31075 & 153.04566 & 161.9840 & 142.2631\\
S/2006 S5 & 22.64912 & 23.53175 & 21.94601 & 0.23184 & 0.38793 & 0.10194 & 167.75079 & 171.5942 & 163.5061\\
S/2006 S7 & 22.90792 & 23.95062 & 22.15544 & 0.45552 & 0.59255 & 0.31550 & 168.84316 & 173.7958 & 162.9456\\
S/2004 S16& 22.90942 & 23.66638 & 22.27512 & 0.15619 & 0.24886 & 0.07616 & 164.67491 & 168.4193 & 160.6690\\
Ymir      & 23.03358 & 24.02542 & 22.30504 & 0.34788 & 0.47584 & 0.22387 & 173.64663 & 177.9365 & 169.0400\\
S/2004 S08& 24.20494 & 25.23716 & 23.41207 & 0.21845 & 0.32804 & 0.11797 & 170.09407 & 173.6581 & 166.2385\\
\\          
\hline      
\end{tabular}
\end{center}
\end{table*}
The outer Solar System is inhabited by different minor bodies populations: comets, Centaurs, trans-neptunian objects (TNO), Trojans and irregular satellites of the giant planets. Apart from comets, whose existence had long been known, the other populations have been discovered in the last century and extensively studied since then. Our comprehension of their dynamical and physical histories has greatly improved (see \cite{sf08a,sf08b} for a general review on the subject) and recent models of the formation and evolution of the Solar System seem to succeed in explaining their evolution and overall structure (see \cite{gom05,mor05,tsi05} for details). There is still no clear consensus, however, on the origin of the irregular satellites. It's widely accepted that their orbital features are not compatible with an \textit{in situ} formation, leading to the conclusion they must be captured bodies. At present, however, no single capture model is universally accepted (see \cite{she06} and \cite{jeh07} for a general review). It is historically believed that the satellite capture occurred prior to the dissipation of the Solar Nebula \citep{pol79}, since the gaseous drag is essential for the energy loss process that leads to the capture of a body in a satellite orbit while crossing the sphere of influence of a giant planet.\\
A direct consequence and implicit assumption of this model is that no major removal of irregular satellites took place after the dissipation of the nebular gas for the actual satellite systems to be representative of the gas-captured ones. If instead this was the case, we are left with two possibilities:
\begin{itemize} 
\item the capture efficiency of the gas--drag mechanism was by far higher than previously assumed in order to have enough satellites surviving to the present time in spite of the removal, or
\item a second phase of capture events, based on different physical mechanisms, took place after the gas dissipation.
\end{itemize}
The same issues apply also to the original formulation of the so called \textit{Pull-Down} scenario \citep{hep77}, also based on the presence of the nebular gas but locating the capture events during the phase of rapid gas accretion and mass growth of the giant planets.\\
Recently, the plausibility of gas-based scenarios has been put on jeopardy.
The comparative study performed by \cite{jes05} pointed out that the giant planets possess similar abundances of irregular satellites, once their apparent magnitudes are scaled and corrected to match the same geocentric distance. Due to the different formation histories of gas and ice giants, the gas-based scenarios cannot supply a convincing explanation to this fact.\\
The \textit{Nice model} \citep{gom05,mor05,tsi05} instead undermines the physical relevance of the gas-based scenarios. Formulated to explain the present orbital structure of the outer Solar System, it postulates that the giant planets formed (or migrated due to the interaction with the nebular gas) in a more compact configuration than the current one. Successively, due to their mutual gravitational interactions, the giant planets evolved through a phase of dynamical rearrangement in which the ice giants and Saturn migrated outward while Jupiter moved slightly inward. The dynamical evolution of the giant planets was stabilised, during this phase, by the interaction with a disc of residual planetesimals. During the migration process, a fraction of the planetesimals can be captured as Trojans \citep{bol08} while the surviving outer planetesimal disk would slowly settle down as the present Kuiper Belt. The perturbations of the migrating planets on the planetesimals would have also caused the onset of the Late Heavy Bombardment on the inner planets. The detailed description of this model and its implications are given in \cite{gom05,mor05,tsi05} and in \cite{mor07}. A major issue with the original formulation of the Nice model was the \textit{ad hoc} choice of the orbits of the giant planets after the dispersal of the nebular gas. Work is ongoing to solve the issue \citep{mor07} by assuming that migration of planets by interaction with the Solar Nebula could lead to a planetary resonant configuration subsequently destroyed by the interaction with the planetesimal disk.\\
The Nice model has two major consequences concerning the origin of the irregular satellites. Due to the large distance from their parent planets and their consequently looser gravitational bounds, no irregular satellites previously captured could have survived the phase of violent rearrangement taking place in the Solar System \citep{tsi05}. In addition, during the rearrangement phase planetesimals where crossing the region of the giant planets with an intensity orders of magnitude superior to the one we observe in the present Solar System \citep{tsi05}, favouring capture by different dynamical processes.\\
In other words, the Nice model doesn't rule out gas-based mechanisms for irregular satellites capture but it implies that they could not have survived till now due to the violent dynamical evolution of the planets. The only exception to the former statement could be represented by the Jovian system of irregular satellites, since Jupiter had a somehow quieter dynamical evolution and its satellites were more strongly tied to the planet due to its intense gravity. For the other giant planets, however, the chaotic orbital evolution created a favourable environment for the capture of new irregular satellites by three body processes (i.e. gravitational interactions during close encounters and collisions). These trapping processes don't significantly depend on the choice of the initial orbital configuration of the planets in the Nice model: they rely mainly on the presence of a residual disk of planetesimals which both stabilises the dynamical evolution of the giant planets and supplies the bodies that can become irregular satellites.\\
In this paper we concentrate on the study of the dynamical features of Saturn's irregular satellites, stimulated by the unprecedented data gathered by the Cassini mission on Phoebe. The overall structure of the satellite system can in fact provide significant clues on its origin and capture mechanism. This dynamical exploration will be the first step to test the feasibility and implications of the collisional capture scenario which was originally proposed by \cite{col71}, which will be the subject of a forthcoming paper.
The work we will present is organised as follows:
\begin{itemize}
\item in section \ref{mean-elements} we will describe the setup we used to simulate the dynamical history of the irregular satellites of Saturn through N-Body integrations and present the proper orbital elements we computed;
\item in section \ref{dynamical-evolution} we will analyse the dynamical evolution of the satellites and search for resonant or chaotic behaviours;
\item in section \ref{collisional-evolution} we will present the updated collisional probabilities between pairs of irregular satellites. We will also estimate the impact probability of bodies orbiting close to Phoebe to explore the relevance of post--capture impacts in sculpting the system;
\item in section \ref{collisional-families} we will apply the Hierarchical Clustering Method described in \cite{zap90,zap94} to search for dynamical families and compare the results with the available observational data;
\item finally, in section \ref{conclusion} we will put together all our results to draw a global picture of Saturn's system of irregular satellites.
\end{itemize}
            
\section{Computation of the mean orbital elements}\label{mean-elements}
            
\begin{figure}
\centering  
\includegraphics[width=8.0cm]{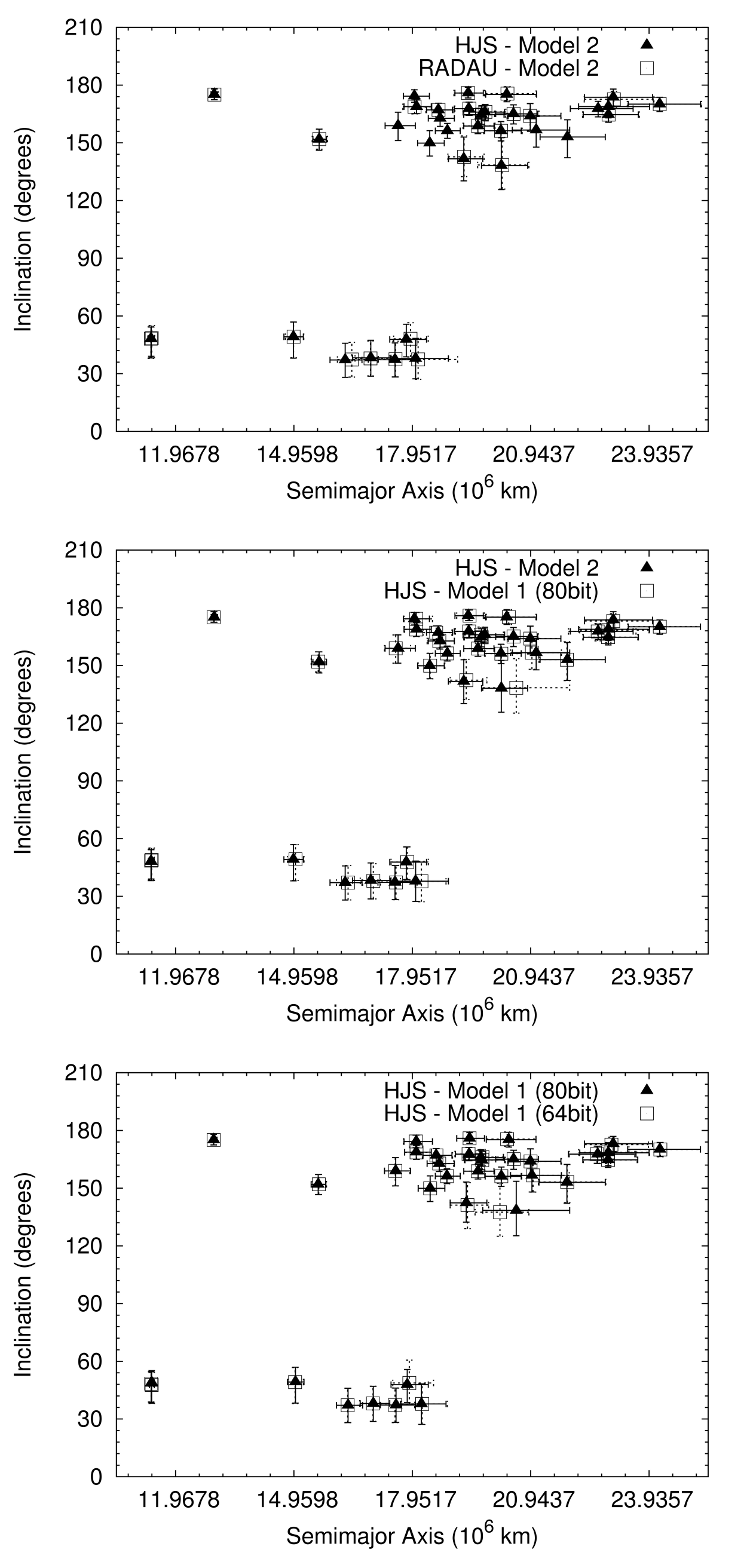}
\caption{Comparison in the $a-i$ plane between the mean orbital elements of Saturn's irregular satellites computed with Model $2$ using HJS and RADAU algorithms (upper plot), Model $1$ and Model $2$ using HJS algorithm (middle plot) and Model $1$ with standard and strict double precision (lower plot). The vertical and horizontal bars show the variation ranges of the elements in the simulations. Distances are expressed in $10^{6}$ km while angles are expressed in degrees. In the simulation with RADAU algorithm we considered only the $26$ irregular satellites known till $2005$.}
\label{model-ai}
\end{figure}
\begin{figure}
\centering
\includegraphics[width=8.0cm]{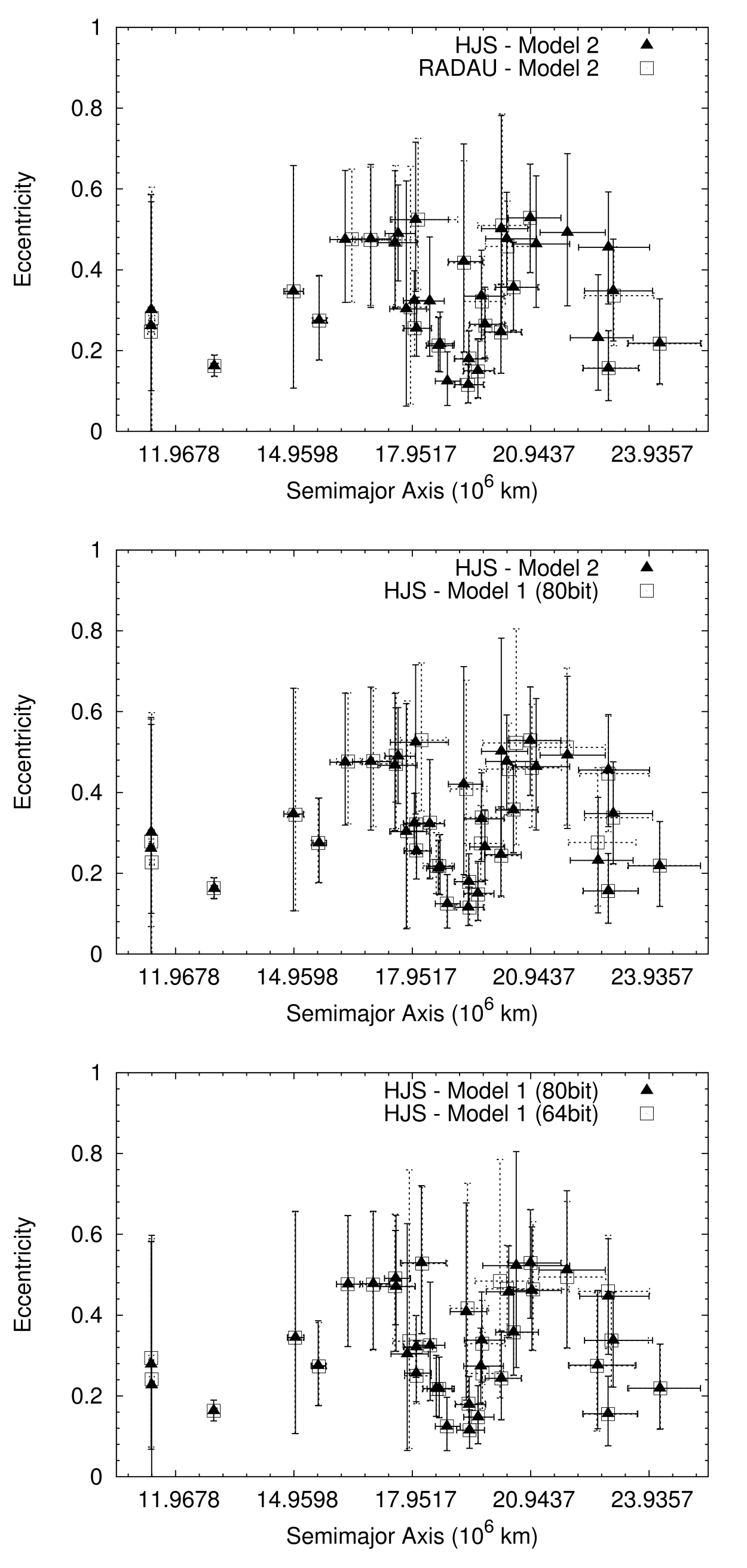}
\caption{Comparison in the $a-e$ plane between the mean orbital elements of Saturn's irregular satellites computed with Model $2$ using HJS and RADAU algorithms (upper plot), Model $1$ and Model $2$ using HJS algorithm (middle plot) and Model $1$ with standard and strict double precision (lower plot). The vertical and horizontal bars show the variation ranges of the elements in the simulations. Distances are expressed in $10^{6}$ km while angles are expressed in degrees. In the simulation with RADAU algorithm we considered only the $26$ irregular satellites known till $2005$.}
\label{model-ae}
\end{figure}
\begin{figure}
\centering
\includegraphics[width=8.0cm]{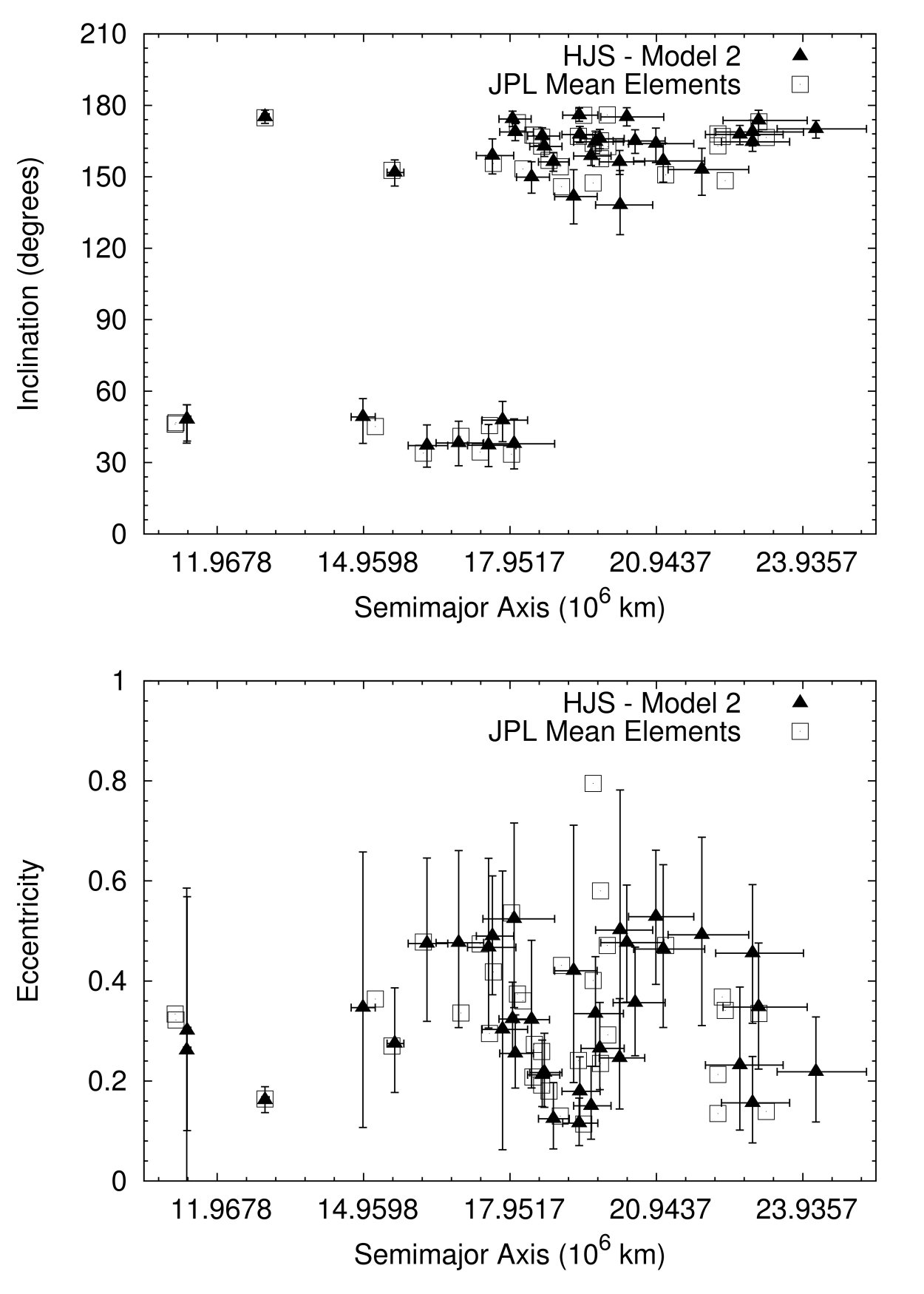}
\caption{Comparison between the mean orbital elements of Saturn's irregular satellites computed with Model $2$ using the HJS algorithm and those from the JPL Solar System Dynamics website. The comparison is shown in the $a-i$ (upper graph) and $a-e$ (lower graph) planes. The vertical and horizontal bars show the variation ranges of the elements in our simulation. Distances are expressed in $10^{6}$ km while angles are expressed in degrees.} 
\label{jpl-comparison}
\end{figure}

Our study of the dynamics of the irregular satellites start from the computations of approximate proper orbital elements. Proper elements may be derived analytically from the nonlinear theory developed by \cite{mek94} or they can be approximated by the mean orbital elements, which can be numerically computed by averaging the osculating orbital elements over a reasonably long time interval which depends on the evolution of the system.\\
To compute the mean orbital elements of Saturn's irregular satellites we adopted the numerical approach and integrated over $10^{8}$ years the evolution of an N--body dynamical system composed of the Sun, the four giant planets and the satellites of Saturn in orbit around the planet. For the satellites, we adopted two different dynamical setups:
\begin{enumerate}
\item Model $1$: $35$ irregular satellites treated as test particles
\item Model $2$: $35$ irregular satellites treated as test particles, in addition Titan and Iapetus treated as massive bodies
\end{enumerate}
Model $1$ basically reproduces the orbital scheme used by other authors in previous works \citep{car02,nes03,cuk04}. Model $2$ includes the perturbing effects of the two outermost regular satellites of the Saturn system. We considered this alternative model to evaluate their contribution to the global stability and to the secular evolution of the irregular satellites.\\
In order to reproduce accurately the orbits of the satellites, the integration timestep has to be less than the shortest period associated to the frequencies of motion of the system: the common prescription for computational celestial mechanics is to use integration timesteps about $1/20$ of the fastest orbital period, usually that of the innermost body. The timesteps we used in our simulations with Models $1$ and $2$ were respectively:
\begin{itemize}
\item $20$ days ($1/22$ of the orbital period of Kiviuq)
\item $0.75$ day ($1/20$ of the orbital period of Titan)
\end{itemize}
The timestep used for Model $2$ greatly limited the length of the simulations by imposing a heavier computational load. A time interval of $10^{8}$ years was a reasonable compromise between the need of accuracy in the computation of proper elements and CPU (i.e. computational time) requirements: this time interval has been used also by \cite{nes03}. We used it for both Models in order to be able to compare their results.\\
The initial osculating orbital elements of the irregular satellites and of the major bodies have been derived respectively from two different ephemeris services:
\begin{itemize}
\item {\bf Natural Satellites} of the \textit{IAU Minor Planet Center}\footnote{\url{http://cfa-www.harvard.edu/iau/NatSats}}
\item {\bf Horizons} of the NASA Jet Propulsion Laboratory\footnote{\url{http://ssd.jpl.nasa.gov/?horizons}}
\end{itemize}
The reference plane in all the simulations was the $J2000$ ecliptic and the initial elements of all bodies referred to the epoch $30$ January $2005$.\\
The numerical simulations have been performed with the public implementation of the HJS (Hierarchical Jacobi Symplectic) algorithm described in \cite{beu03} and based on the SWIFT code \citep{led94}. The symplectic mapping scheme of HJS is applicable to any hierarchical 
system without any \textit{a priori} restriction on the orbital structure. The public implementation does not include the effects of planetary oblateness or tidal forces which are not considered in our models\footnote{Preliminary tests on a timespan of several $10^{6}$ years, performed with a new library (on development) which allows the treatment of such effects in HJS algorithm, show no major discrepancies in the computed mean orbital elements and no qualitative differences on the chaos measures we will describe in section \ref{dynamical-evolution}, thus supporting the analysis and the results reported in the present paper.}. The HJS algorithm proved itself a valuable tool to study the dynamical evolution of satellite systems since its symplectic scheme supports multiple orbital centres. It easily allows to include in the simulations both Titan and Iapetus which were not taken into account in the previous works on irregular satellites.\\
To test the reliability of HJS, we performed an additional run based on Model $2$ using the RADAU algorithm \citep{rad85} incorporated in the MERCURY 6.2 package by J.E. Chambers \citep{cha99}. Due to the high computational load of RADAU algorithm compared to symplectic mapping, we limited this simulation to the $26$ irregular satellites known at the end of $2005$ for Saturn, which means we excluded S/2004 S19 and the S/2006 satellites. Even with this setup, RADAU's run required about $6$ months of computational time: as a comparison, HJS's run based on full Model $2$ took approximately a month.\\
The averaged (proper) orbital elements computed in Models $1$ and $2$ are given in tables \ref{proper-elements1} and \ref{proper-elements2}, together with the minimum and maximum values reached during the simulations. The same elements are visually displayed in fig. \ref{model-ai} and \ref{model-ae}.
At first sight, the mean elements obtained with Models $1$ and $2$ appear to be approximately the same. However, there are some relevant differences between the two sets: this is the case of Tarvos and S/2004 S18 and of Kiviuq and Ijiraq, whose radial ordering results inverted. The perturbations by Titan and Iapetus significantly affect the secular evolution of the satellite system. We will show in section \ref{dynamical-evolution} that the changes are more profund that those shortly described here.\\
The differences in the mean elements computed with RADAU and HJS on the same model (Model $2$) are shown in the top panels of fig. \ref{model-ai} and \ref{model-ae} and are significantly less important: they appear to be due to the presence of chaos in the system. Finally, in fig. \ref{jpl-comparison} we confronted the mean elements computed from Model $2$ with the average orbital elements available on the JPL Solar System Dynamics website\footnote{\url{http://ssd.jpl.nasa.gov/?sat_elem}}. In this case the differences between the two dataset are more marked, probably depending on the different integration--averaging time used to compute them.\\
In the following discussions, when not stated differently, we will always be implicitly referring to the mean elements computed with Model $2$ since it better represents the real dynamical system. We would like to stress that the validity of our mean elements is proved over $10^{8}$ years and only through the analysis of the satellites' secular evolution we can be able to assess if they can be meaningful on longer timescales.\\
As a final remark, the two major gaps in the radial distribution of Saturn's irregular satellites, the first centred at Phoebe and extending from $11.22 \times 10^{6}$ km to $14.96 \times 10^{6}$ km and the second located between $20.94 \times 10^{6}$ km and $22.44 \times 10^{6}$ km, will be discussed respectively in section \ref{collisional-evolution} and \ref{dynamical-evolution}.
\begin{figure}
\centering
\includegraphics[width=8.0cm]{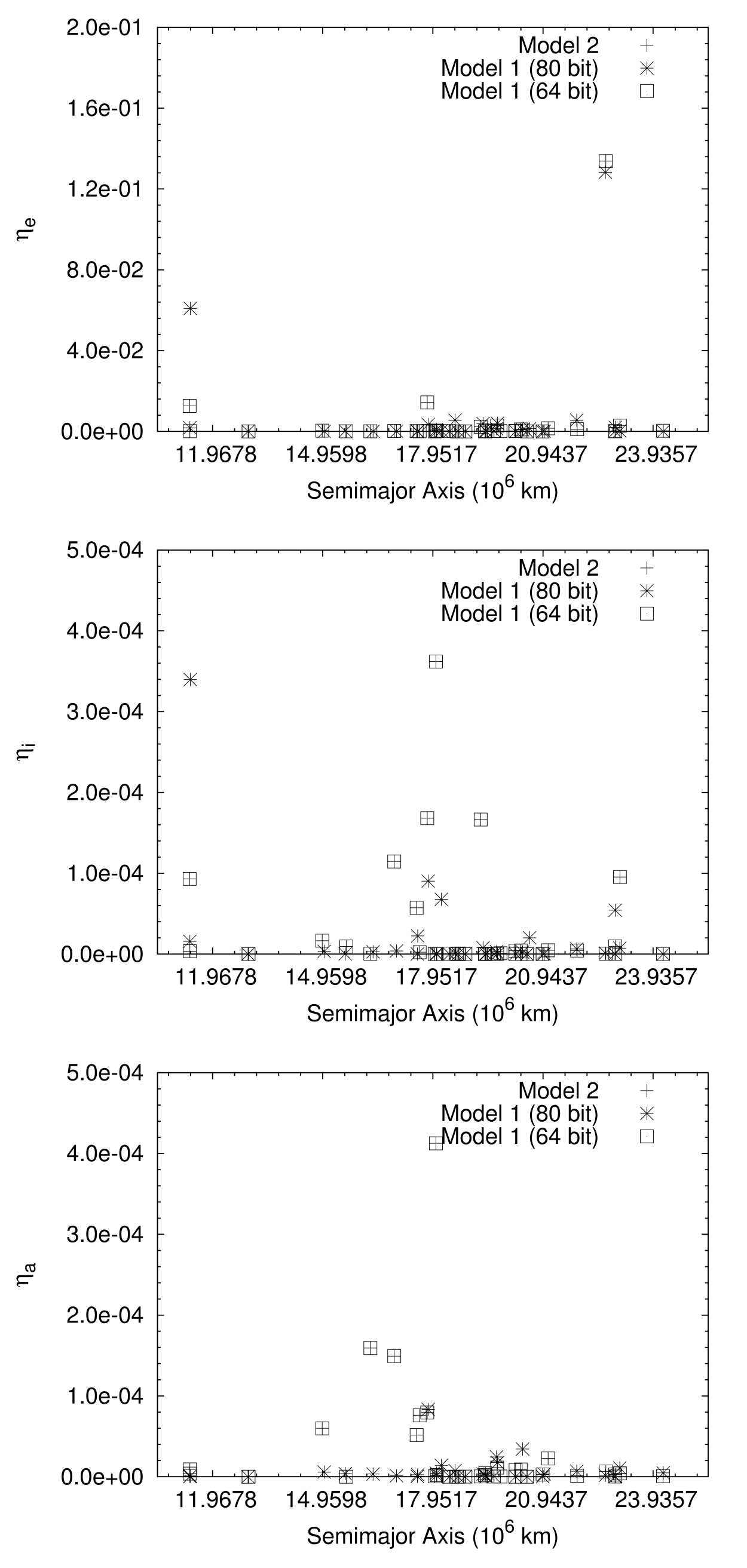}
\caption{Values of the $\eta$ parameters for the irregular satellites of Saturn. From top to bottom the plots show the $\eta$ values for eccentricity, inclination and semimajor axis respectively. The results concern Models $1$ and $2$ and the additional simulation based on Model $1$ where strict double precision ($64$ bit) was adopted in the computations instead of the standard extended precision ($80$ bit). The change in the numerical precision has major effects on various satellites which, as a consequence, evolve on chaotic orbits.}
\label{etas-sat}
\end{figure}

\FloatBarrier
\section{Evaluation of the dynamical evolution}\label{dynamical-evolution}
\begin{figure*}
\centering
\includegraphics[width=16.0cm]{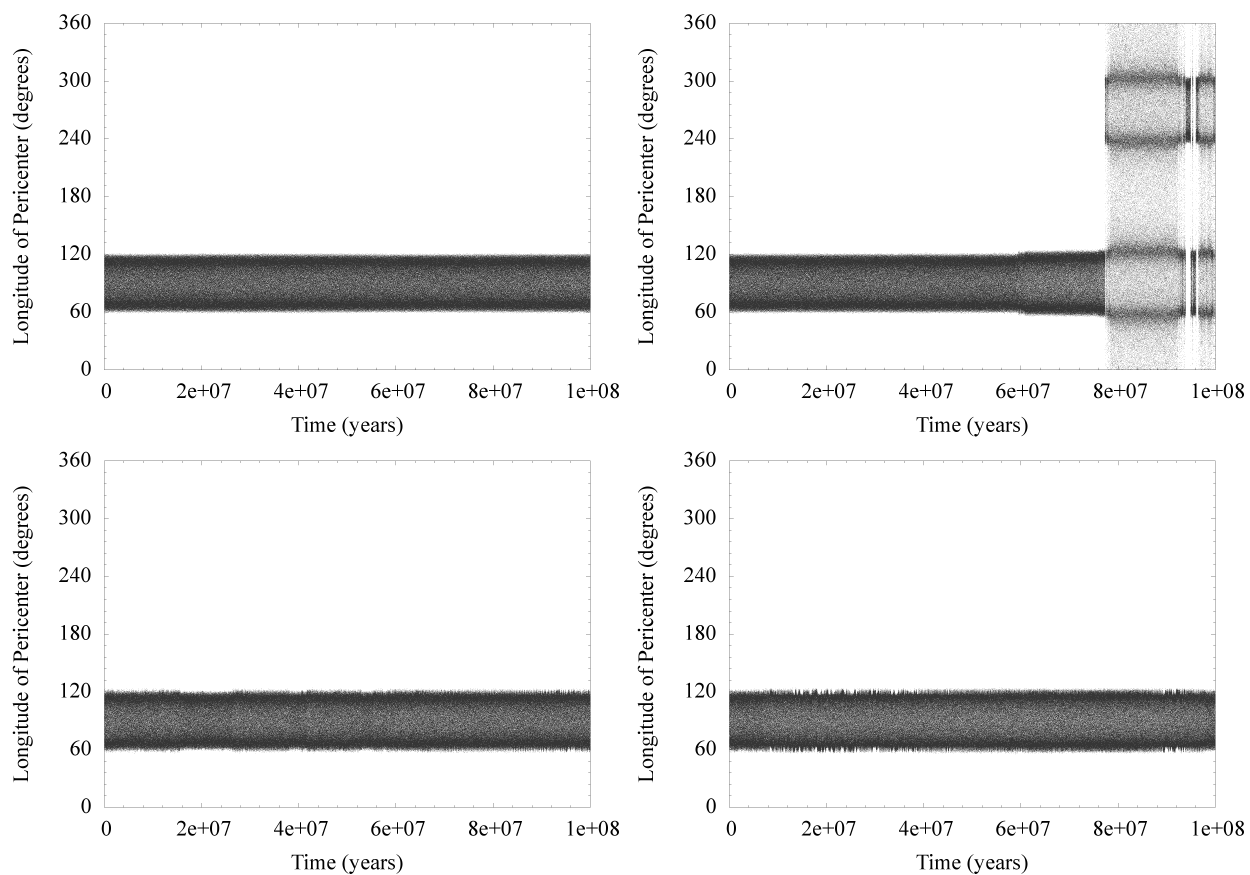}
\caption{Secular evolution of Ijiraq's longitude of pericenter in (from top left, clockwise direction) Model $2$ computed with HJS algorithm, Model $2$ computed with RADAU algorithm, Model $1$, Model $1$ with strict ($64$ bit) double precision. Angles are expressed in degrees and time in years. In all cases computed with HJS code, Ijiraq's longitude of pericenter is into a stable librational regime, indicating that the satellite is locked into a Kozai resonance with the Sun. In the case computed with RADAU, the satellite is initially trapped into the same resonance but it breaks the resonant regime during the last $2 \times 10^{7}$ years of the simulation. Escape from the resonance is not permanent, since the satellite is temporarily captured at least other two times before the end of the simulation. Ijiraq's behaviour is probably due to the perturbations of Titan and Iapetus.}
\label{ijiraq-peri}
\end{figure*}
\begin{figure}
\centering
\includegraphics[width=8.0cm]{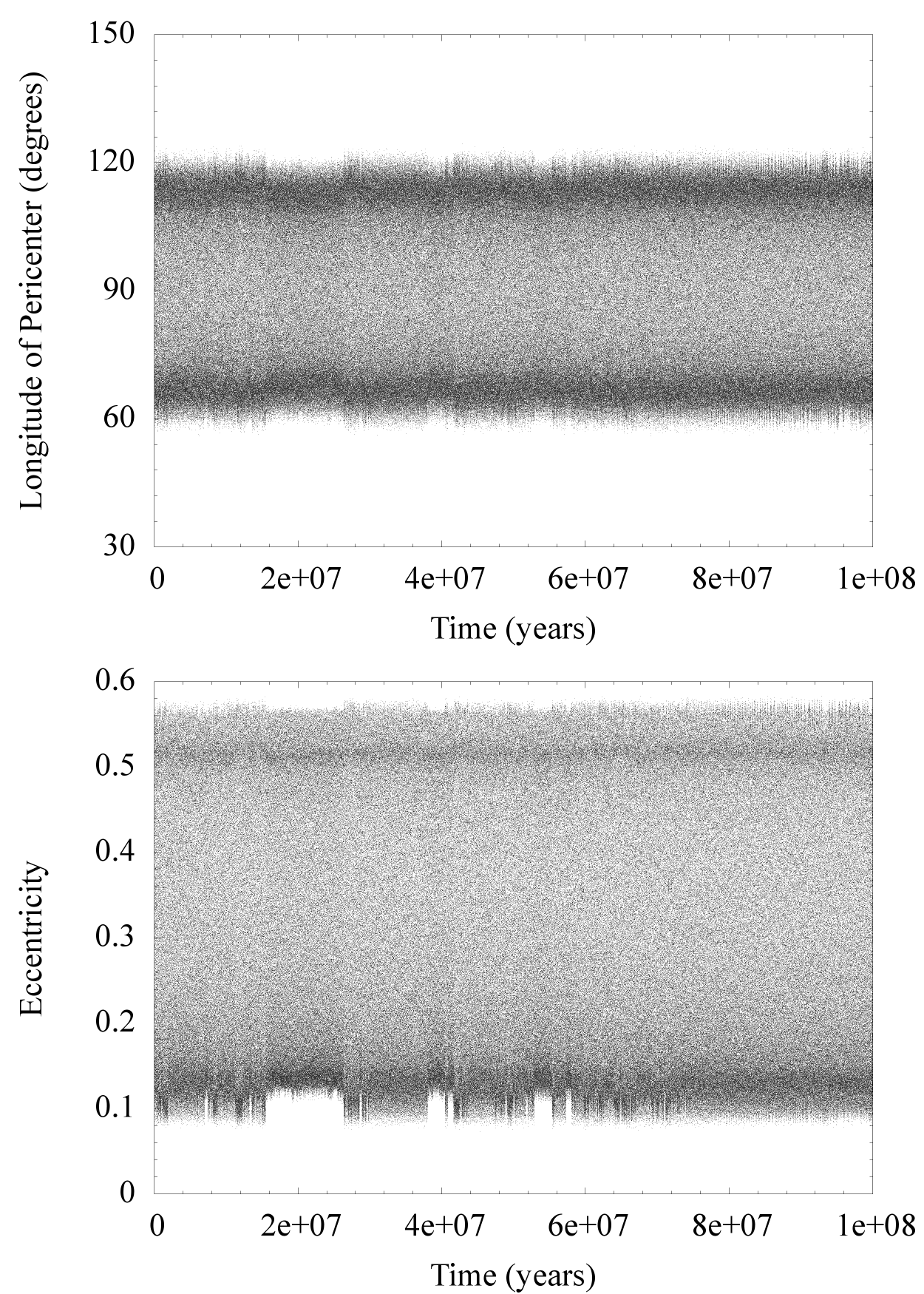}
\caption{Evolution of Ijiraq's longitude of pericenter and eccentricity in Model $1$ simulated with HJS algorithm using strict ($64$ bit) double precision. The longitude of pericenter and the eccentricity have limited non-periodic changes which may be ascribed to the perturbations of Jupiter since Titan and Iapetus are not included in the model.}
\label{ijiraq-pe}
\end{figure}
\begin{figure}
\centering
\includegraphics[width=8.0cm]{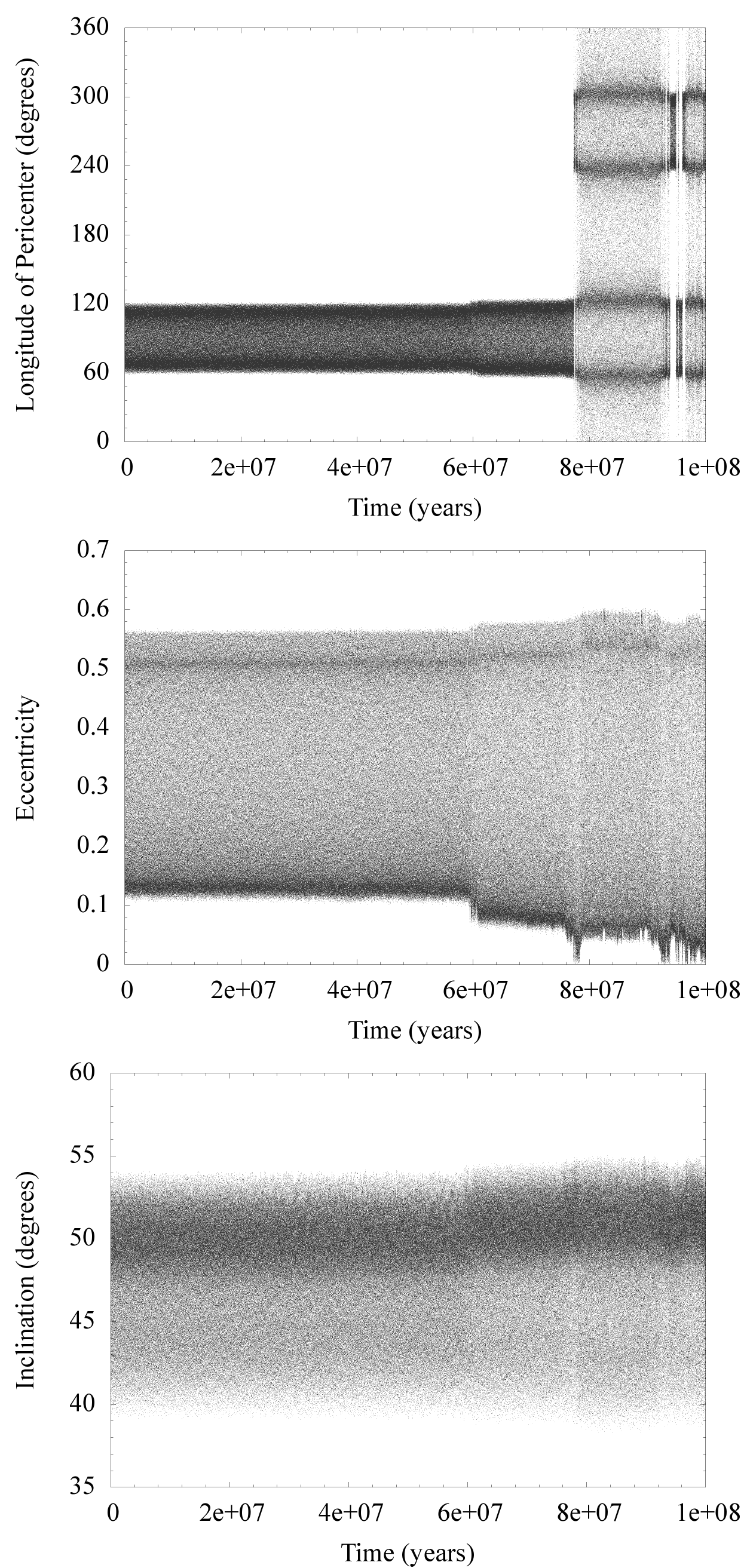}
\caption{Evolution of Ijiraq's longitude of pericenter, eccentricity and inclination (from top to bottom) in Model $2$ integrated with RADAU. The dynamical evolution of the satellite has a sudden change around $6 \times 10^{7}$ years leading to a progressive increase in the eccentricity oscillations. Also the libration amplitude slightly grows until the longitude of pericenter starts to circulate at $8 \times 10^{7}$ years. Since fig. \ref{ijiraq-pe} proved that the effects of the giant planets are quite limited after the onset of the Kozai resonance with the Sun, Titan and Iapetus are responsible for the perturbed evolution of the satellite.}
\label{ijiraq-pei}
\end{figure}

\begin{figure*}
\centering
\includegraphics[width=16.0cm]{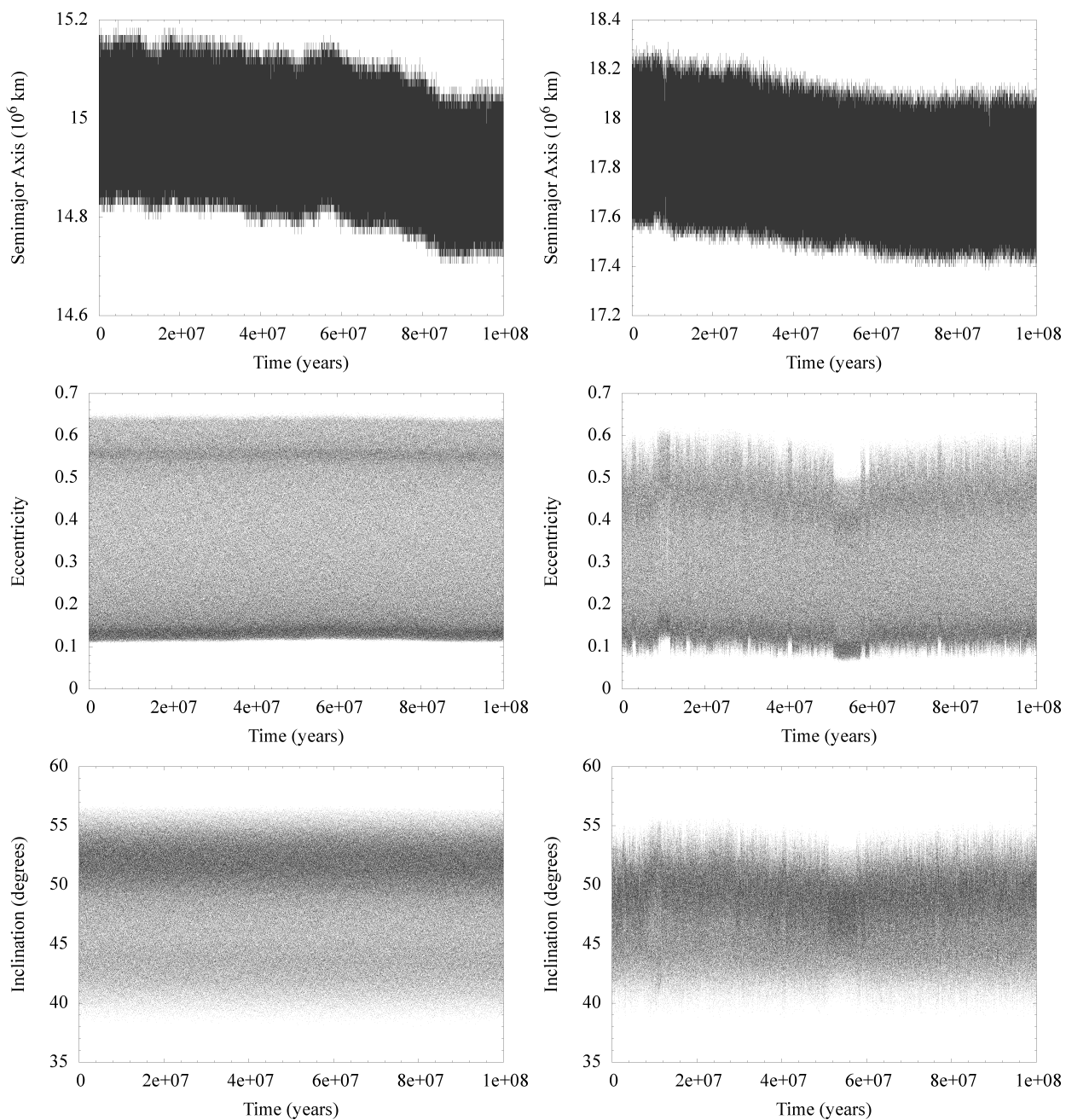}
\caption{Comparison between the evolution of semimajor axis, eccentricity and inclination (from top to bottom) of Paaliaq (left column) and Siarnaq (right column) in Model $2$ integrated with HJS. Both satellites suffer a slight inward displacement, limited to about $\sim1-2\%$, yet Paaliaq's eccentricity and inclination show a regular behaviour whereas Siarnaq's ones have a coupled variation at about $1 \times 10^{7}$ years and $5 \times 10^{7}$ years. This behaviour is reproduced in both Models (see fig. \ref{siarnaq-et} and \ref{siarnaq-it} for further details on Siarnaq's evolution).}
\label{paaliaq-siarnaq}
\end{figure*}

\begin{figure*}
\centering
\includegraphics[width=16cm]{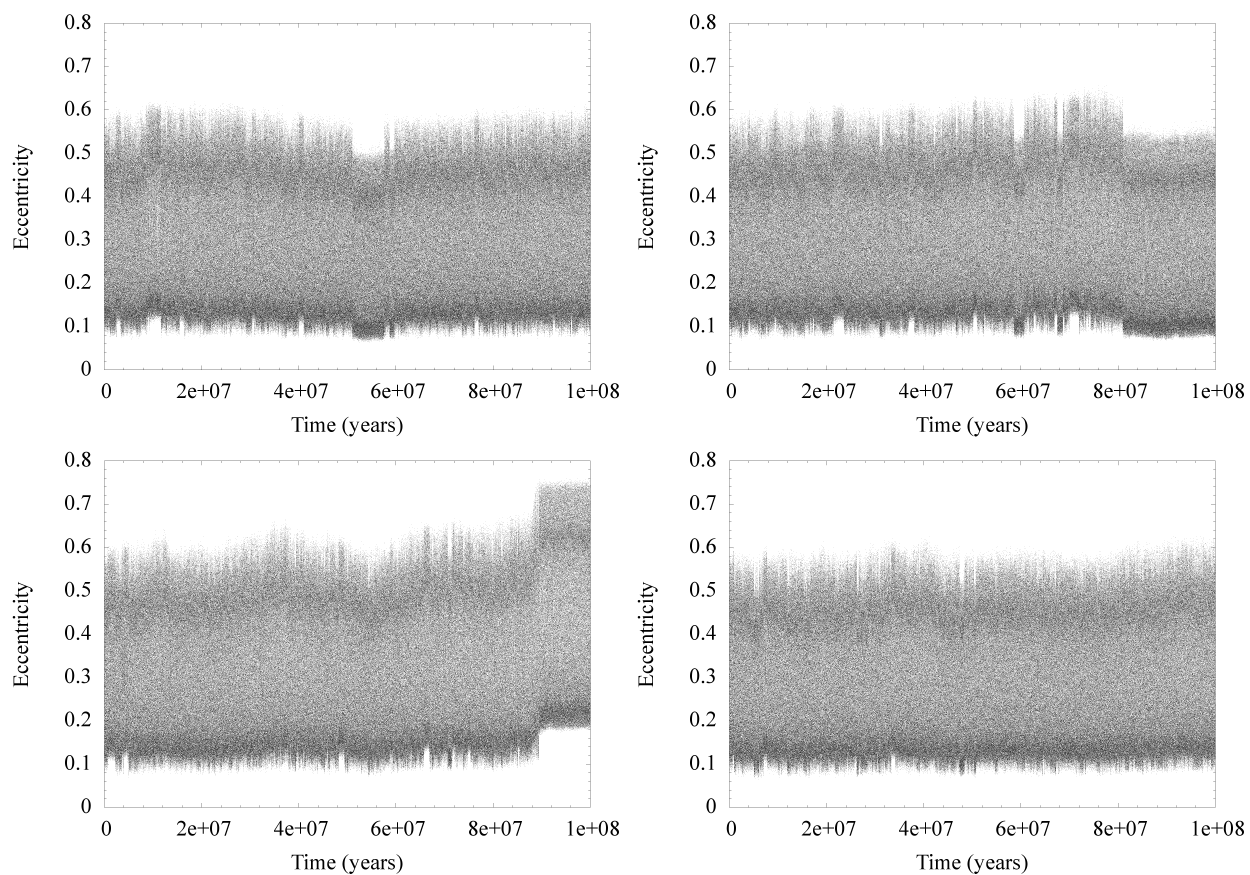}
\caption{Secular evolution of the eccentricity of Siarnaq. From top left, clockwise direction, the plots show the outcome of Model $2$ - HJS algorithm, Model $2$ - RADAU algorithm, Model $1$ - HJS with standard precision and Model $1$ - HJS with strict ($64$ bits) double precision. Siarnaq's eccentricity clearly shows evidence of chaotic evolution in Model $2$ and Model $1$ ($64$ bit). In Model $1$ ($80$ bit) Siarnaq's eccentricity appears quasi-stationary at all timescales.}
\label{siarnaq-et}
\end{figure*}

\begin{figure*}
\centering
\includegraphics[width=16cm]{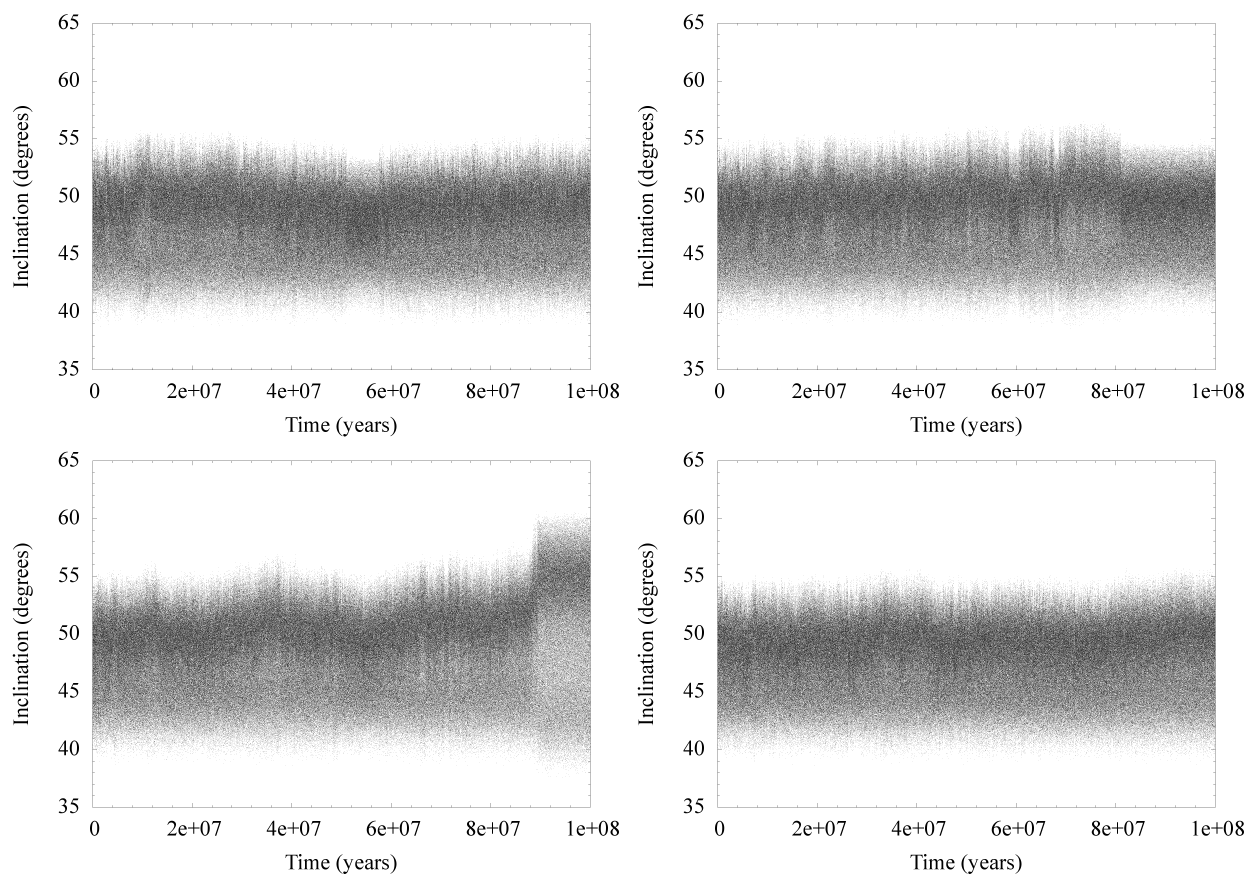}
\caption{Secular evolution of the inclination of Siarnaq. From top left, clockwise direction, the plots show the outcome of Model $2$ - HJS algorithm, Model $2$ - RADAU algorithm, Model $1$ - HJS with standard precision and Model $1$ - HJS with strict ($64$ bits) double precision. While Paaliaq radial displacement is coupled to a regular evolution of the other orbital elements (see fig. \ref{paaliaq-siarnaq}), Siarnaq's eccentricity (see fig. \ref{siarnaq-et}) and inclination show clearly the presence of chaotic features in Model $2$ and Model $1$ ($64$ bit). In Model $1$ ($80$ bit) Siarnaq's inclination appears quasi-stationary at all timescales.}
\label{siarnaq-it}
\end{figure*}

\begin{figure*}
\centering
\includegraphics[width=16.0cm]{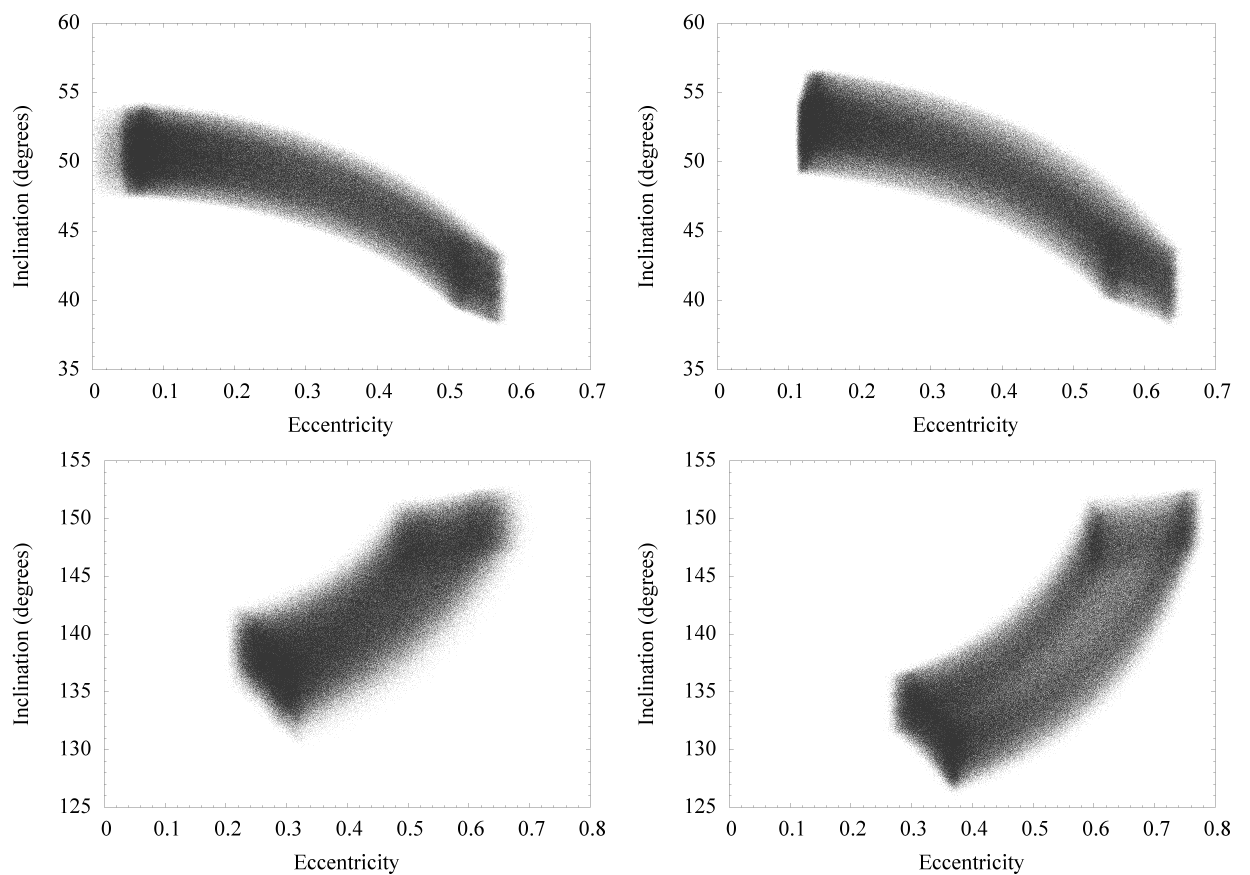}
\caption{Anticorrelation of inclination and eccentricity (see also fig. \ref{kozai-ei}) in the dynamical evolution of (clockwise from top left) Ijiraq, Paaliaq, Bestla (S/2004 S18) and Narvi. These features, observed in both Models $1$ and $2$, might imply that these satellites were prevented from entering the Kozai cycle with the Sun by the combined gravitational perturbations of the other giant planets. It is noteworthy that the only satellites in a full Kozai regime are the two innermost ones, for which the external perturbations are less effective.}
\label{kozai-arc}
\end{figure*}

\begin{figure*}
\centering
\includegraphics[width=16.0cm]{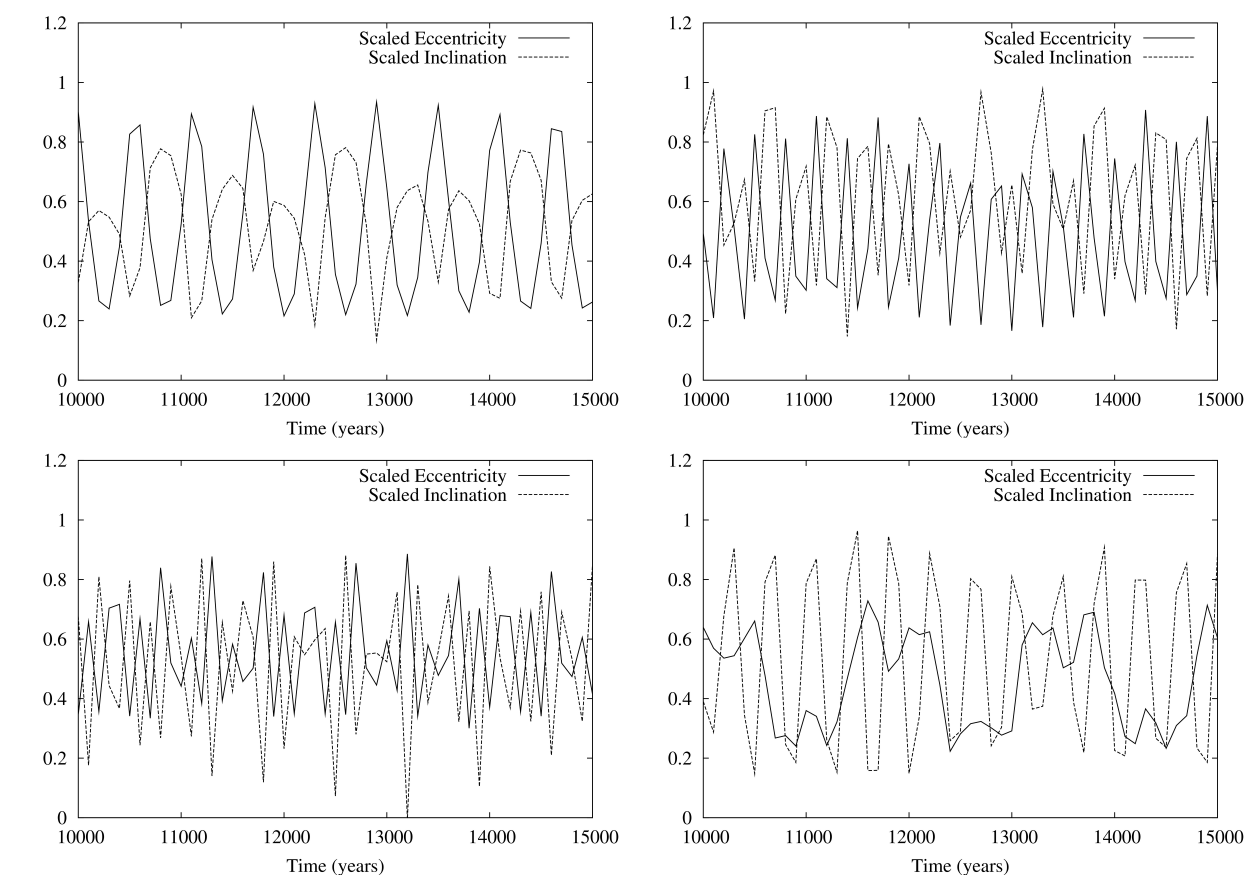}
\caption{Phase displacement of inclination and eccentricity in the dynamical evolution of (clockwise from top left) Ijiraq, Paaliaq, Bestla (S/2004 S18) and Narvi. Both eccentricity and inclination have been rescaled and shifted to show variations of the same order of magnitude. The plots referring to prograde satellites show the relation between eccentricity and inclination, those referring to retrograde satellites show the relation between eccentricity and the inclination supplemental angle (i.e. $180^\circ - i$). Bestla's evolution show alternating phases of correlation and anticorrelation between eccentricity and inclination: anticorrelation phases generally last twice than correlation phases. This effect is at the basis of Bestla's thicker arc in fig. \ref{kozai-arc}. The evolutions of Bestla's inclination and eccentricity are strongly coupled (see fig. \ref{forced-e} for details) and this coupling interferes with solar perturbations causing Kozai regime.}
\label{kozai-ei}
\end{figure*}

\begin{figure*}
\centering
\includegraphics[width=16.0cm]{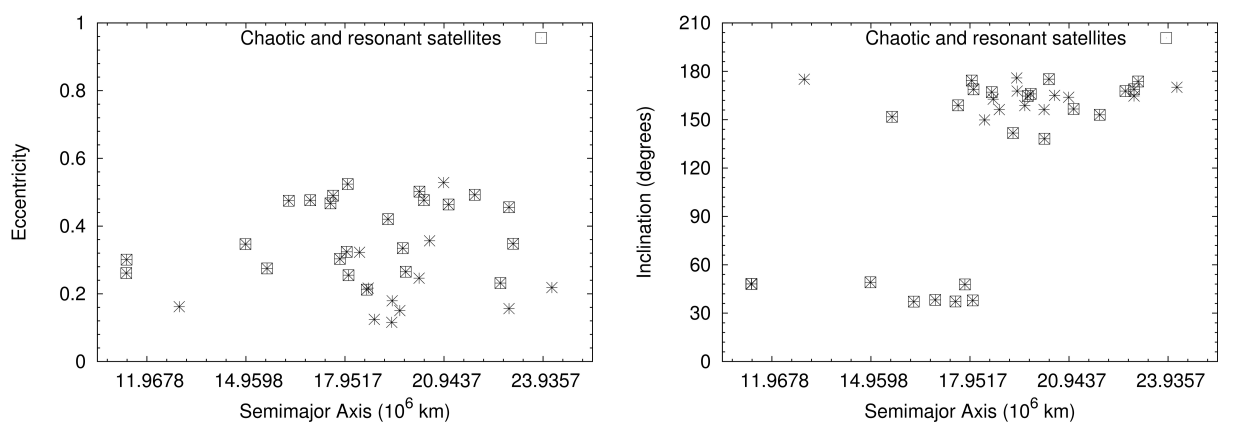}
\caption{Chaotic and resonant irregular satellites of Saturn's system. The square symbols identifies the irregular satellites showing resonant or chaotic features in their dynamical evolution. The data are presented in the $a-e$ (left plot) and $a-i$ (right plot) planes.}
\label{chaotic-satellites}
\end{figure*}

\begin{figure*}
\centering
\includegraphics[width=16cm]{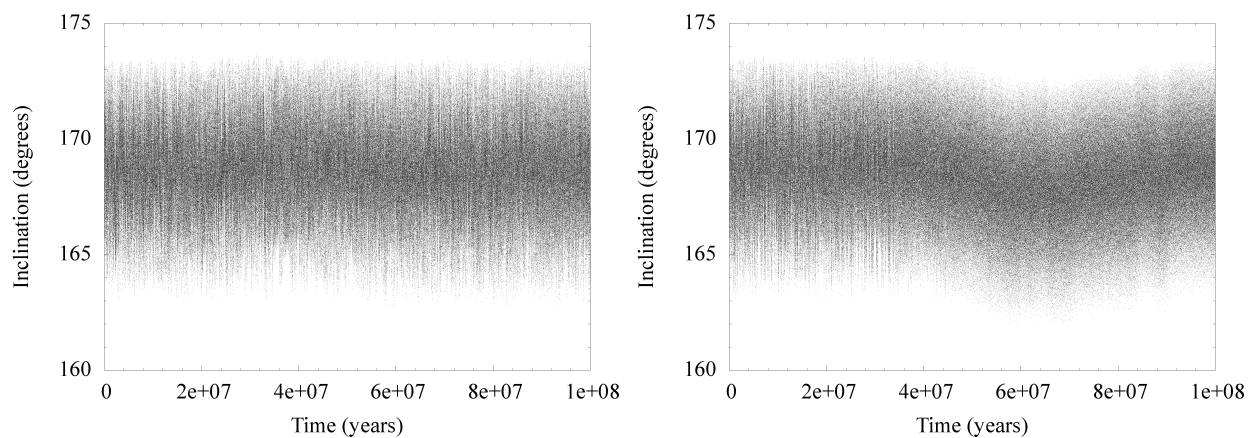}
\caption{Secular evolution of the inclination of S/2006 S7 in Model $2$ (left panel) and Model $1$ computed with strict ($64$ bits) double precision (right panel). While the first case show a regular behaviour, the last one presents an evident chaotic jump in inclination. This jump may be related to the entry of the satellite into a chaotic region. This event does not occur in the Model $1$ computed with standard precision, probably due to the different numerical setup.}
\label{2006S7-it}
\end{figure*}

\begin{figure*}
\centering
\includegraphics[width=16cm]{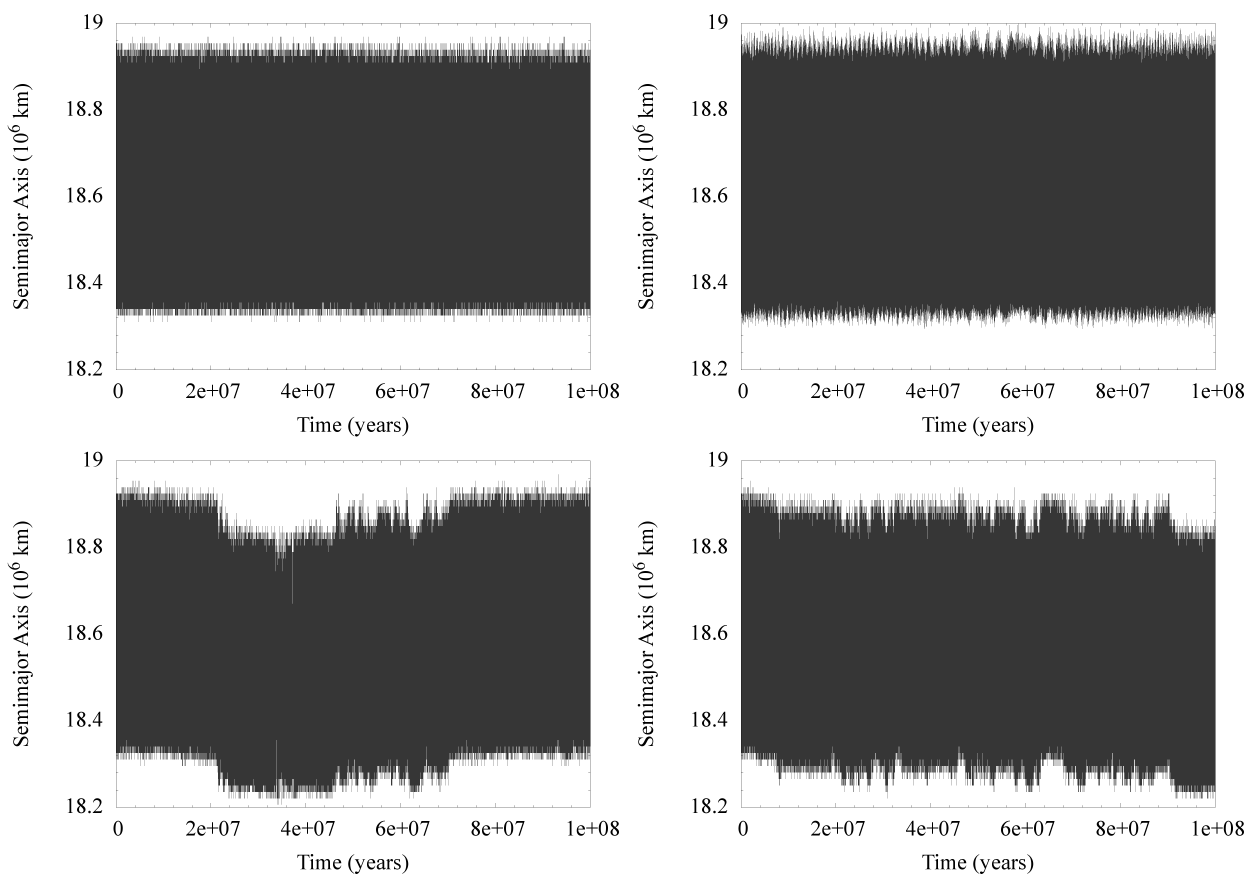}
\caption{Secular evolution of the semimajor axis of Mundilfari. From top left, clockwise direction, the plots show the outcome of Model $2$ - HJS algorithm, Model $2$ - RADAU algorithm, Model $1$ - HJS with standard precision and Model $1$ - HJS with strict ($64$ bits) double precision. The simulations based on Model $2$ show a regular behaviour on the whole timespan, independently of the employed algorithm. Both cases based on Model $1$ show instead an irregular evolution diagnostic of chaotic behaviour. Titan and Iapetus act like orbital stabilisers for Mundilfari.}
\label{mundilfari-at}
\end{figure*}

\begin{figure*}
\centering
\includegraphics[width=16cm]{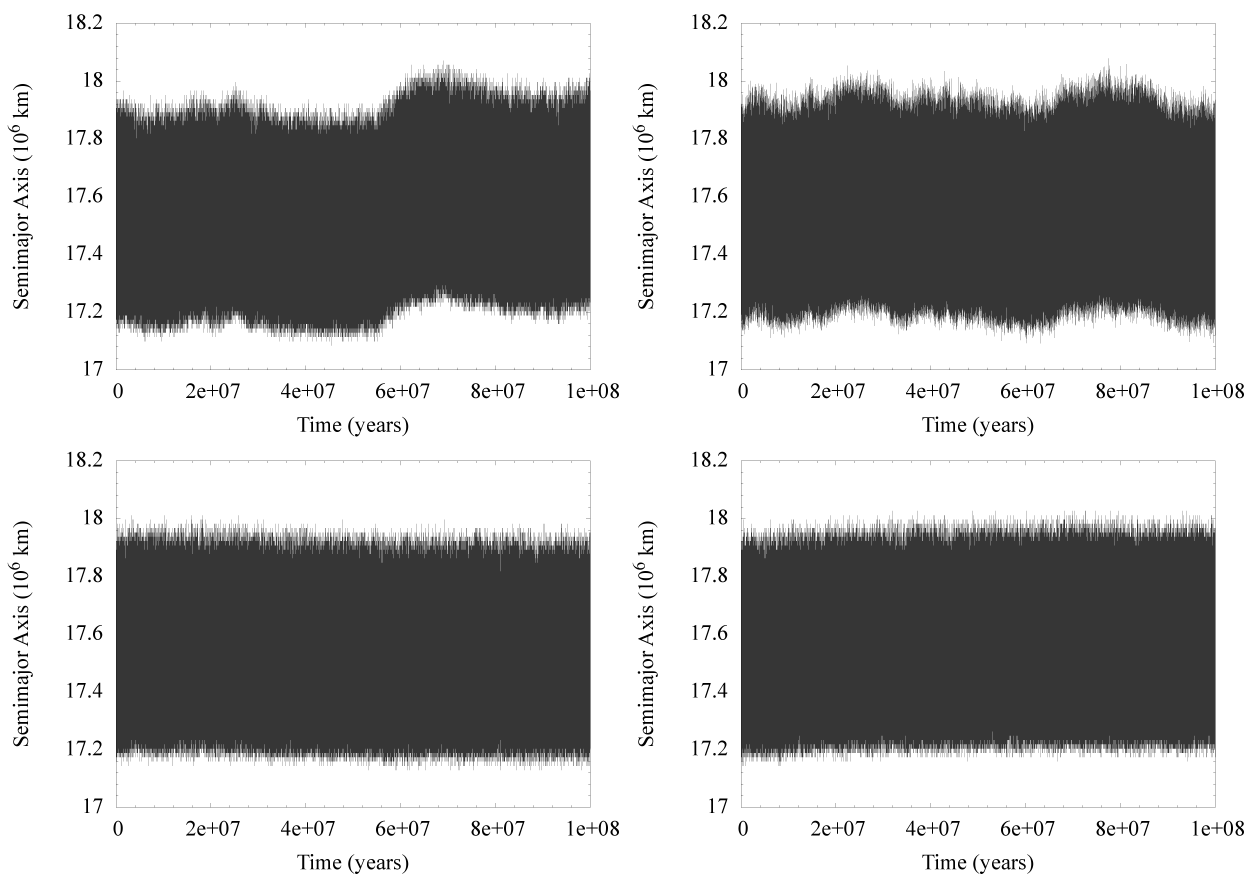}
\caption{Secular evolution of the semimajor axis of Erriapo. From top left, clockwise direction, the plots show the outcome of Model $2$ - HJS algorithm, Model $2$ - RADAU algorithm, Model $1$ - HJS with standard precision and Model $1$ - HJS with strict ($64$ bits) double precision. Contrary to the case showed in fig. \ref{mundilfari-at}, Titan and Iapetus cause an irregular evolution of Erriapo. In Model $1$ the satellite has a regular behaviour.}
\label{erriapo-at}
\end{figure*}

\begin{figure*}
\centering
\includegraphics[width=16cm]{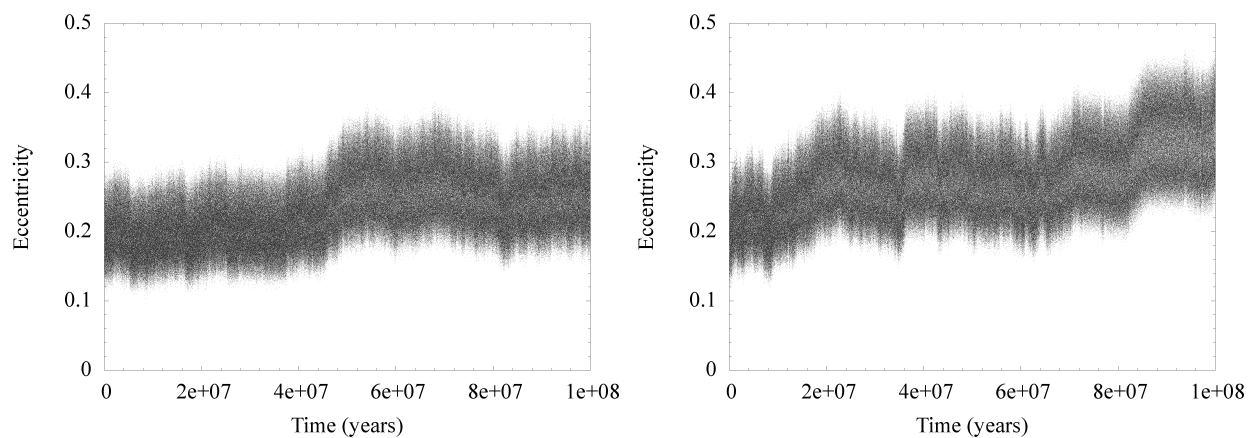}
\caption{Secular evolution of the eccentricity of S/2006 S5 in Model $2$ computed with HJS (left panel) and Model $1$ computed using standard precision (right panel). The chaotic nature of the satellite orbital motion can be easily inferred by the secular behaviour of its eccentricity. Similar conclusions can be drawn from the inspection of semimajor axis and inclination.}
\label{2006S5-et}
\end{figure*}

\begin{figure*}
\centering
\includegraphics[width=16.0cm]{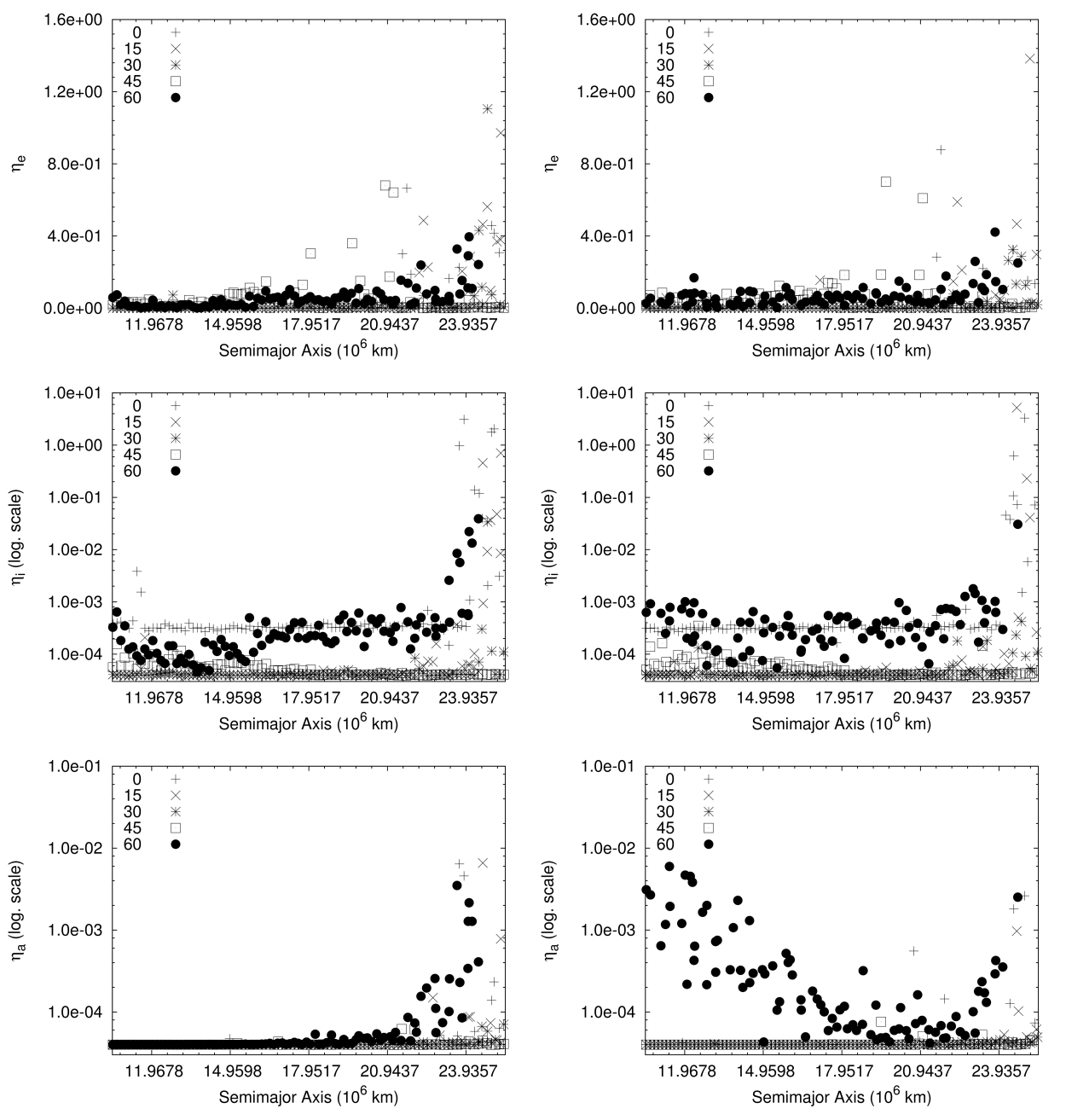}
\caption{Evolution of the $\eta$ parameters with semimajor axis for prograde test particles in Model $1$ (left column) and Model $2$ (right column). From top to bottom, the plots show the values of $\eta_e$, $\eta_i$, $\eta_a$ respectively. Values of $\eta$ near $0$ indicate quasi periodic motion, higher values indicate increasingly chaotic or resonant behaviour. Distances are expressed in $10^6$ km.}
\label{pro-eta}
\end{figure*}

\begin{figure*}
\centering
\includegraphics[width=16.0cm]{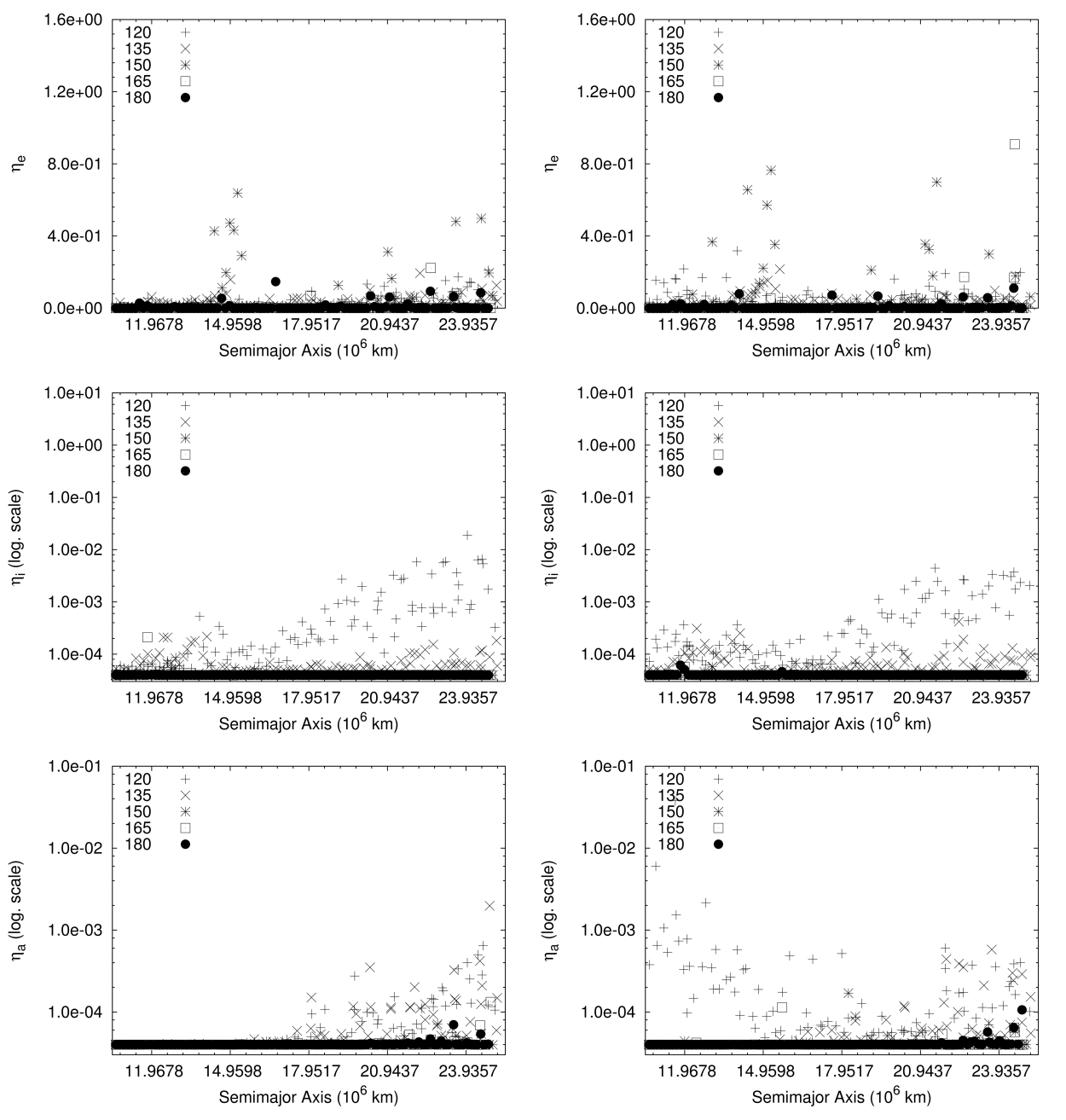}
\caption{Evolution of the $\eta$ parameters with semimajor axis for retrograde test particles in Model $1$ (left column) and Model $2$ (right column). From top to bottom, the plots show the values of $\eta_e$, $\eta_i$, $\eta_a$ respectively. Values of $\eta$ near $0$ indicate quasi periodic motion, higher values indicate increasingly chaotic or resonant behaviour. Distances are expressed in $10^6$ km.}
\label{retro-eta}
\end{figure*}

\begin{figure*}
\centering
\includegraphics[width=16.0cm]{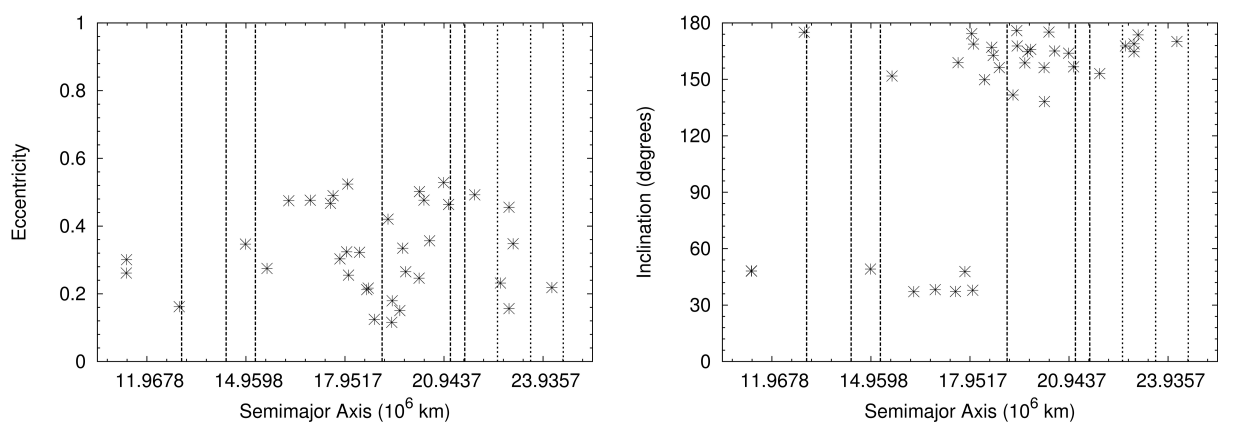}
\caption{Location of the irregular satellites in the $a-e$ (upper plot) and $a-i$ (lower plot) planes respect to the unstable regions identified by the simulations computing the evolution of regularly sampled test particles. The unstable regions, perturbed by Jupiter, are marked by dashed and dotted lines. Dotted lines indicate regions which are excited for all values of inclination. Dashed lines are related to those regions which get perturbed for high absolute inclination values (i.e. $i \sim 150^{\circ}$).}
\label{excited}
\end{figure*}

\begin{figure*}
\centering
\includegraphics[width=16.0cm]{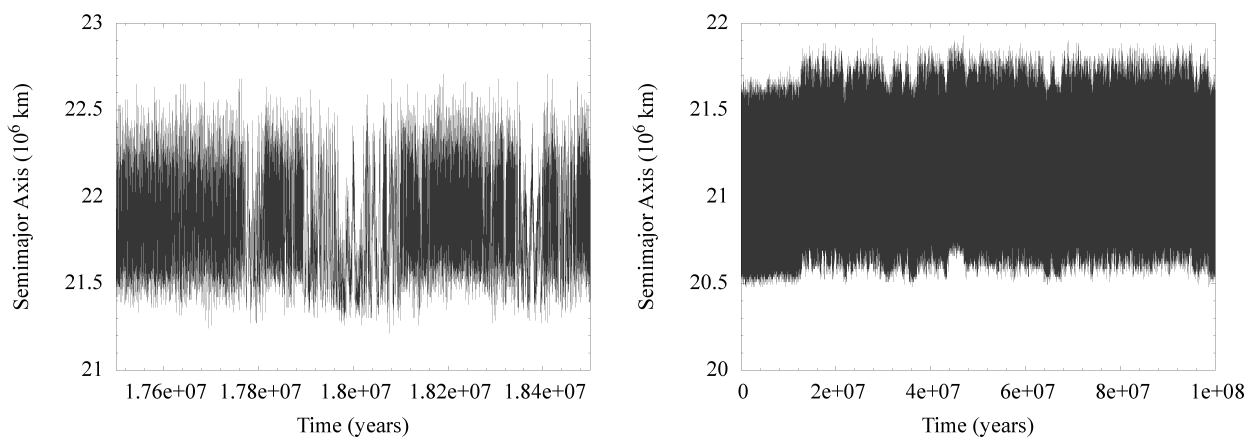}
\caption{Secular evolution of the semimajor axis of S/2006 S2 Kari (detail, left graph) and S/2006 S3 (detail, right graph) in model $2$. While on different timescales and with different intensities, both the satellites have a chaotic evolution possibly due to Jupiter's perturbations through the Great Inequality with Saturn.}
\label{chaos-gap}
\end{figure*}

The mean orbital elements given in the previous section give a global view of the dynamical evolution of the satellite system. However, to better understand the features of the mean elements and of the differences observed between the different models we need to have a better insight on the individual secular evolution of the satellites. We start our analysis by looking for hints of chaotic or resonant behaviour. We adopt a variant of the mean actions criterion (see \cite{mor02} and references therein) introduced by \cite{cuk04} to study the dynamics of irregular satellites.
This modified criterion is based on the computation of the $\eta$ parameter, defined as
\begin{equation}\label{eq-eta1}
\eta_e = \sum_{j=1}^{10} \, \frac{\left( \left\langle e \right\rangle_{j} - \left\langle e \right\rangle\right)^{2}}{\left\langle e \right\rangle^{2}}
\end{equation} 
where
\begin{equation}\label{eq-eta2}
\left\langle e \right\rangle_{j} = \frac{10}{T}\int^{j\left( T/10 \right)}_{\left( j-1\right) \left( T/10 \right)} e\,dt
\end{equation}
and
\begin{equation}\label{eq-eta3}
\left\langle e \right\rangle = \frac{1}{T}\int^{T}_{0} e\,dt.
\end{equation}
The $\eta$ parameter is a measure of the temporal uniformity of the secular evolution of the eccentricity. As shown in eq. \ref{eq-eta1}, \ref{eq-eta2} and \ref{eq-eta3}, its value is estimated by summing the differences between the mean values $\left\langle e \right\rangle_j$ of eccentricity computed on a fixed number of time intervals ($10$ in equations \ref{eq-eta1} and \ref{eq-eta2}) and the mean value $\left\langle e \right\rangle$ on the whole timespan covered by the simulations. If the motion of the considered body is regular or quasi-periodic, the mean value of eccentricity should change little in time and the $\eta$ parameter should approach zero. On the contrary, if the motion is resonant or chaotic, the mean value of eccentricity would depend on the time interval over which it is computed and, therefore, $\eta$ would assume increasing values. Obviously, the reliability of this method depends on the relationship between the number and extension of the time intervals considered and the frequencies of motion: if the timescale of the variations is shorter than the length of each time interval, the averaging process can mask the irregular behaviour of the motion and output an $\eta$ value lower than the one really describing the satellite dynamics. On the contrary, a satellite having a regular behaviour but whose orbital elements oscillate with a period longer than the considered time intervals may spuriously appear chaotic. There is no {\it a priori} prescription concerning the number of time intervals to be used: after a set of numerical experiments, we opted for the value ($10$ time intervals) suggested in the original work by \cite{cuk04}.\\
In the following we extend the idea of \cite{cuk04} by applying a similar analysis also to inclination and semimajor axis, thus introducing $\eta_a$ and $\eta_i$ as:
\begin{equation}
\eta_a = \sum_{j=1}^{10} \, \frac{\left( \left\langle a \right\rangle_{j} - \left\langle a \right\rangle\right)^{2}}{\left\langle a \right\rangle^{2}}
\end{equation}
and
\begin{equation}
\eta_i = \sum_{j=1}^{10} \, \frac{\left( \left\langle i \right\rangle_{j} - \left\langle i \right\rangle\right)^{2}}{\left\langle i \right\rangle^{2}}
\end{equation}
with corresponding definitions of the quantities involved.\\
We present the $\eta$ values in fig. \ref{etas-sat}, where we compare the output of Model $1$, of Model $2$ and of an alternative version of Model $1$ where a different precision in the floating point calculations have been used. We we will shortly explain the reason of this duplication.\\
The coefficient showing the largest variations is $\eta_e$ (see fig. \ref{etas-sat}, top panel) and the extent of the dynamical variations for each satellite can be roughly evaluated by the square root of $\eta_e$. These variations are on average around $5-10\%$ with peak values of about $30\%$. The corresponding variations in semimajor axis and inclination, as derived from the $\eta_a$ and $\eta_i$, are on average more limited ranging from $1\%$ to $2\%$. Since the values of $\eta_e$ are not significantly correlated to those of $\eta_a$ and $\eta_i$, our guess is that the different coefficients are indicators of different dynamical effects.\\
We used a Model $1$ characterised by a different numerical precision in the calculations to unmask the presence of chaos in the system. The standard set-up on $x86$ machines for double precision computing is the so called \textit{extended precision}, where the floating point numbers are stored in double precision variables ($64$ bit) but a higher precision ($80$ bit) is employed during the computations. This is also the set-up we originally used in Model $1$, which in the graphs is labelled ``Model $1$ ($80$ bit)''. We ran an additional simulation forcing the computer to strictly use double precision, thus employing $64$ bits also during the computations. The results of this simulation are labelled as ``Model $1$ ($64$ bit)''. Hereinafter, unless differently stated, when we refer generically to Model $1$ we intend the $80$ bit case. From a numerical point of view, the change in computing precision should in principle affect only the last few digits of each floating point number, since the values are stored in $64$ bit variables in both cases. Regular motions should be slightly affected by the change while chaotic motions should show divergent trajectories. By comparing the different values of the $\eta$ parameters for ``Model $1$ ($64$ bit)'' and ``Model $1$ ($80$ bit)'' we find some satellites with drastically different values, a clear indication of chaotic motion.\\
\begin{figure*}
\centering
\includegraphics[width=16.0cm]{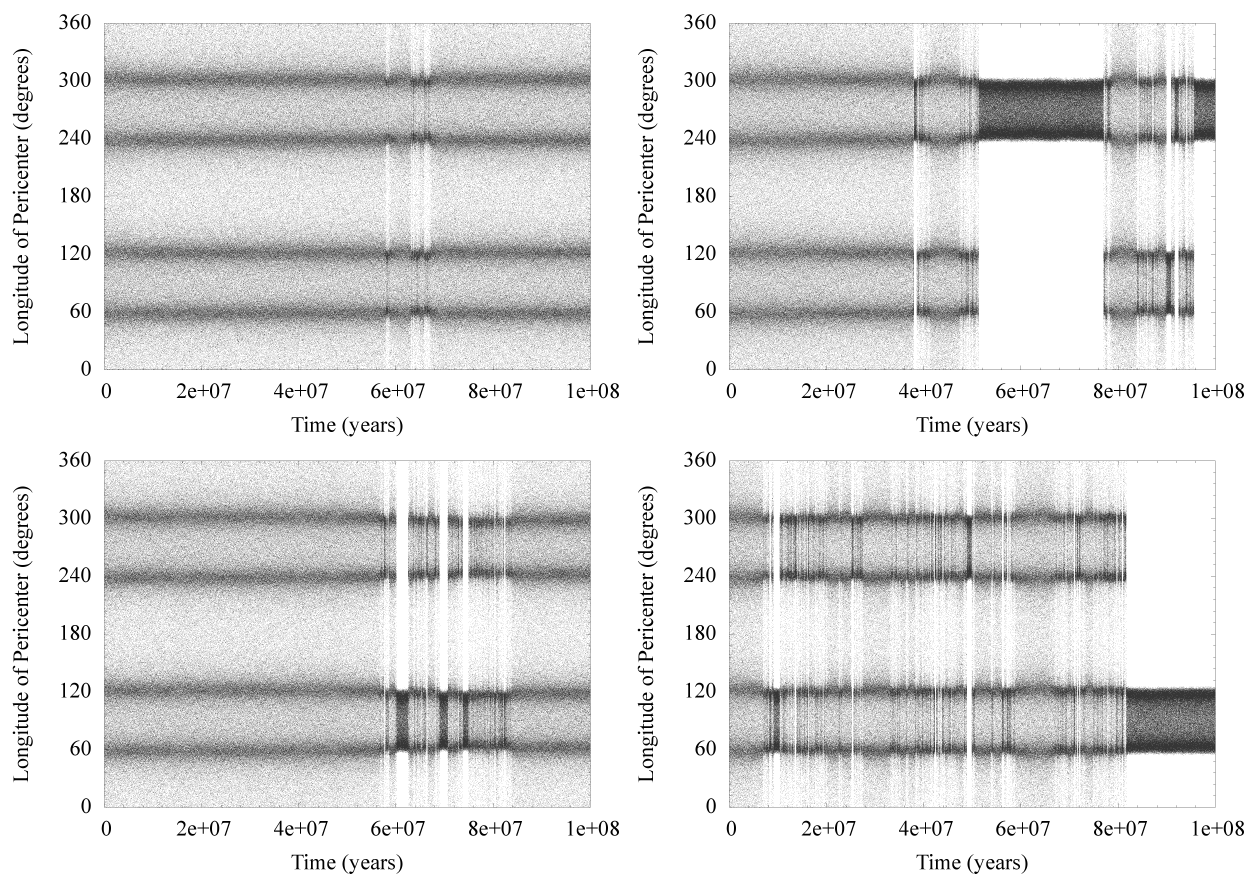}
\caption{Secular evolution of Kiviuq's longitude of pericenter. From top left in clockwise direction we plot the outcome of Model $2$ computed with HJS algorithm, Model $2$ computed with RADAU algorithm, Model $1$, Model $1$ with strict ($64$ bit) double precision. Angles are expressed in degrees and time in years. Transitions between circulation and libration appear in both the dynamical models and independently from the numerical codes. Librations in Model $2$ - HJS occur between $6 \times 10^{7}$ and $7 \times 10^{7}$ years and are extremely short--lived (the angle mainly circulates). The librations are also evenly distributed around $90^{\circ}$ and $270^{\circ}$. The same model computed with RADAU shows long--lived librational periods centred at $270^{\circ}$. Also in Model $1$ the librations are long--lived but they concentrated around $90^{\circ}$. In Model $1$ (bottom right panel) at the end of the simulation the satellite enters a libration phase which lasts about $2 \times 10^7$ years. The differences between the four cases argue for the chaotic nature of Kiviuq's dynamical evolution. Titan and Iapetus intervene altering the dynamical history of the satellite.}
\label{kiviuq-peri}
\end{figure*}
After this preliminary global study of the satellite system, we have performed a more detailed analysis of selected objects whose behaviours have been already investigated in previous papers. \cite{car02}, \cite{nes03} and \cite{cuk04} reported four cases of resonant motion between the irregular satellites of Saturn:
\begin{itemize}
\item Kiviuq and Ijiraq being in the Kozai regime with the Sun;
\item Siarnaq and Paaliaq, whose pericenters appeared tidally locked to the Sun.
\end{itemize}
The authors computed the orbital evolution of these satellites with a modified version of Swift N--Body code, where the planets were integrated in the heliocentric reference frame while the irregular satellites in planetocentric frames. This dynamical structure is similar to that of our Model $1$. We extracted from our simulations the data concerning the same objects studied in \cite{car02}, \cite{nes03} and \cite{cuk04} and we analysed their behaviour.
The motion of Ijiraq as from our Models $1$ and $2$ integrated with HJS (see fig. \ref{ijiraq-peri}, first three panels from top left in counterclockwise direction) appears to evolve in a stable Kozai regime with the Sun. The longitude of pericenter librates around $90^{\circ}$ with an amplitude of $\pm30^{\circ}$. The simulations based on Model $2$ and Model $1$ ($80$ bit precision) show a regular secular behaviour of the orbital elements while the simulation based on Model $1$ ($64$ bit precision) shows periods in which the range of variation of the eccentricity shrinks by a few percent in correspondence to a similar behaviour of the longitude of pericenter (see fig. \ref{ijiraq-pe} for details). The output obtained from Model $2$ with RADAU algorithm (fig. \ref{ijiraq-pei}) is instead significantly different. Ijiraq is in a stable Kozai regime for the first $6 \times 10^{7}$ years, then it experiences a change in the secular behaviour of both eccentricity and inclination and, finally, the longitude of pericenter $\varpi$ circulates for about $1.3 \times 10^{7}$ years. From then on, $\varpi$ alternates phases of circulation and libration around both $90^{\circ}$ and $270^{\circ}$.\\
Kiviuq shows an even more complex behaviour (see fig. \ref{kiviuq-peri}). All the simulations (Models $1$ and $2$ and both HJS and RADAU algorithms) predict transitions between periods of circulation and libration of $\varpi$. The centre of libration changes depending on the numerical algorithm: the secular evolution computed with RADAU algorithm show a prevalence of librational phases around $270^{\circ}$. In the case of HJS algorithm the dominant libration mode is around $90^{\circ}$. While the libration cycles were all characterised by an amplitude of about $\pm30^{\circ}$, their durations vary from case to case. In Model $2$ computed with HJS $\varpi$ mostly circulates with only a few short--lived librational periods. In Model $1$ ($80$ bit) the satellite enters the Kozai regime with the Sun only after $8 \times 10^{7}$ years, while Model $1$ ($64$ bit) shows a behaviour more similar to that of Model $2$, even if with longer--lived librational phases. Model $2$ computed with RADAU shows a behaviour similar to that of Model $1$. As for Ijiraq, such different dynamical evolutions confirm the chaotic behaviour of the satellite orbit.\\
Ijiraq and Kiviuq are examples of a different balancing between the perturbations acting on the trajectories of these satellites. Ijiraq appears dominated by the Kozai resonance caused by the Sun's influence while Titan, Iapetus and the other planets play only a minor role. On the opposite side, Kiviuq is significantly perturbed by Titan, Iapetus and the other planets which prevent its settling into a stable Kozai regime with the Sun and cause its chaotic evolution.\\
For Siarnaq and Paaliaq our updated dynamical model do not confirm the resonant motions found in previous publications. In presence of Titan and Iapetus (Model $2$) both the satellites show evidence of a limited but systematic variation of their semimajor axes. They monotonically migrate inwards by $\sim2\%$ during the timespan covered by our simulations (see fig. \ref{paaliaq-siarnaq}). Paaliaq has a regular evolution of the eccentricity and inclination (fig. \ref{paaliaq-siarnaq}, left column), Siarnaq (fig. \ref{paaliaq-siarnaq}, right column) during its radial migration presents phases of chaotic variations of both eccentricity and inclination (see fig. \ref{siarnaq-et} and \ref{siarnaq-it} for details).
We cannot rule out major changes in semimajor axis (possibly periodic?) on longer timescales. Paaliaq, during its orbital evolution, might enter more dynamically perturbed regions (see fig. \ref{phoebe-gap} for further details).\\
The comparison of our results with previously published papers suggest that the three--body approximation adopted in previous analytical works was not accurate enough to be used as a reference model for the dynamical evolution of Saturn's irregular satellites. An additional feature arguing against a simplified three--body approximation is the following one. When we illustrate the secular evolution of the satellites in the $e-i$ plane, in several cases we obtained a thick arc (see fig. \ref{kozai-arc} for details) having opposite orientation for prograde and retrograde satellites. It indicates a global anticorrelation of the two orbital elements (see fig. \ref{kozai-ei} for details) and it appears more frequently among the satellites integrated with Model $2$.
This anticorrelation of eccentricity and inclination is a characteristic feature of the Kozai regime. The arc--like feature is in fact manifest in the two previously discussed cases: Kiviuq and Ijiraq. By inspecting the dynamical histories of all the other satellites, we found the following with the same feature: Paaliaq, Skathi, Albiorix, Erriapo, Siarnaq, Tarvos, Narvi, Bebhionn (S/2004 S11), Bestla (S/2004 S18), Hyrokkin (S/2004 S19). Three of these satellites have an inclination close to the critical one leading to Kozai cycles. Farbauti (S/2004 S9), Kari (S/2006 S2), S/2006 S3 and Surtur (S/2006 S7) show a more dispersed arc--like feature suggesting that external gravitational perturbations by the massive satellites and planets interfere with the Kozai regime due to solar gravitational effects. This hypothesis is confirmed by the fact that in a few cases the arc--like feature is present only in Model $2$ were Titan and Iapetus are included.\\
Additional differences in the chaotic evolution of some satellites can be ascribed to the presence of Titan and Iapetus.
According to our simulations $23$ satellites show a chaotic behaviour. Aside from the previously mentioned Kiviuq, Ijiraq, Paaliaq and Siarnaq, the list is completed by: Albiorix, Erriapo, Tarvos, Narvi, Thrymr, Ymir, Mundilfari, Skathi, S/2004 S10, Bebhionn (S/2004 S11), S/2004 S12, S/2004 S13, S/2004 S18, S/2006 S2, S/2006 S3, S/2004 S4, S/2006 S5, S/2006 S7 and S/2006 S8 (see fig. \ref{chaotic-satellites}). Some of them are chaotic only in one of the two dynamical Models (e.g. fig. \ref{2006S7-it}, fig. \ref{mundilfari-at} and fig. \ref{erriapo-at}) while others in both the Models (see fig. \ref{2006S5-et} as an example). The most
appealing interpretation is that Titan and Iapetus perturb the dynamical evolution of the irregular satellites either by stabilising otherwise chaotic orbits (fig. \ref{mundilfari-at}) or causing chaos (fig. \ref{erriapo-at}). Another interesting feature is that in most cases the chaotic features appeared in the secular evolution of the semimajor axis with eccentricity and inclination exhibiting a regular, quasi--periodic evolution. As a consequence, the present structure of the outer Saturnian system could be not representative of the primordial one. We will further explore this issue in section \ref{collisional-families}.\\
The presence of chaos in the dynamical evolution of the irregular satellites appears to be the major driver of the differences between the outcome of Model $1$ ($80$ and $64$ bit) and $2$ (HJS and MERCURY RADAU algorithms). The chaotic nature of the trajectories is probably also at the origin of the differences (e.g. the alternation between resonant and not resonant phases in Kiviuq's evolution) we noticed between our simulations based on Model $1$ and those published in the literature by \cite{car02}, \cite{nes03} and \cite{cuk04}, which were based on a similar dynamical scheme. The interplay between the inclusion of Titan's and Iapetus' gravitational perturbations and the presence of chaos can finally be invoked to explain the differences between the results obtained with Models $1$ and $2$.\\
To explore in more detail the effects of Jupiter, Titan and Iapetus on the dynamics of irregular satellites we performed two additional sets of simulations where we sampled with a large number of test particles the phase space populated by the satellites. We distributed $100$ test particles in between $10.47 \times 10^{6}$ km and $25.43 \times 10^{6}$ km on initially circular orbits and integrated their trajectories for $10^{6}$ years. $10$ simulations have been performed each with a different value of the initial inclination of the particles. We considered $i = 0^{\circ},\,15^{\circ},\,30^{\circ},\,45^{\circ},\,60^{\circ}$ for prograde orbits, and $i=120^{\circ},\,135^{\circ},\,150^{\circ},\,165^{\circ},\,180^{\circ}$ for retrograde orbits.
To compute the orbital evolution we used the HJS N--Body code and the dynamical structure of Models $1$ and $2$. The initial conditions for the massive perturbing bodies (Sun, giant planets and major satellites) and the timesteps were the same used in our previous simulations.\\
To analyse the output data we looked at the mean elements and the $\eta$ parameters. In fig. \ref{pro-eta} and \ref{retro-eta} we show some examples of the outcome where the influence of Titan and Iapetus is manifest while comparing the outcome of Model $1$ vs. Model $2$. For particles in the inclination range $[30^{\circ}-60^{\circ}]$ and $[120^{\circ}-150^{\circ}]$ the values of the $\eta_e$ are significantly different. In addition, we verified the leading role of Jupiter in perturbing the system. We computed the test particle orbits switching off the gravitational attraction of the planet. The dynamically perturbed regions shown fig. \ref{excited} disappear or are reduced (Titan and Iapetus are still effective) when Jupiter's pull is switched off.\\
By inspecting the mean orbital elements of irregular satellites in fig. \ref{excited} one may notice as those orbiting close to the perturbed regions have a ``stratification'' in eccentricity typical of a resonant population. It is possible that the dynamical features we observe today are not primordial but a consequence of a secular interaction with Jupiter. One of these features is the \textit{gap} between about $20.94 \times 10^{6}$ km and $22.44 \times 10^{6}$ km in the radial distribution of the mean elements of the irregular satellites. This gap was particularly evident till June $2006$, since no known satellite populated that region. With the recent announcement of $9$ additional members of Saturn's irregular satellites, two objects are now known to cross this region: S/2006 S2 and S/2006 S3. Both these bodies have their mean semimajor axis falling near to perturbed regions. According to our simulations, the semimajor axis of both satellites show irregular jumps (see fig. \ref{chaos-gap}) typical of chaotic behaviours. S/2006 S2 has phases of regular evolution alternated with periods of chaos lasting from a few $10^{4}$ up to about $10^{5}$ years (see fig. \ref{chaos-gap}, left plot). In the case of S/2006 S3, we observed a few large jumps in semimajor axis over the timespan covered by our simulations (see fig. \ref{chaos-gap}, right plot). The evolutions of both the eccentricity and the inclination of the satellites appear more regular, with long period modulations and beats.\\
We applied the frequency analysis in order to identify the source of the perturbations leading to the chaotic evolution of the two satellites. By analysing the $h$ and $k$ non singular variables defined with respect to the planet, we found among the various frequencies two with period of $900$ years and $1800$ year respectively. These values are close to the Great Inequality period ($\sim883$ years) of the almost resonance betwen Jupiter and Saturn. This is possibly one source of the chaotic behaviour of the satellites. In addition, the inspection of the upper plot in fig. \ref{retro-eta} shows that the $\eta_a$ parameters of the test particles populating this radial region increase when Titan and Iapetus are included. By comparing the frequencies of motion of the two irregular satellites with those of Titan and Iapetus we find an additional commensurability.
The frequencies $7.1 \times 10^{-4}~yr^{-1}$ and $7.4 \times 10^{-4}~yr^{-1}$ that are present in the power spectrum of S/2006 S2 and S/2006 S6, respectively, are about twice the frequency $3.4 \times 10^{-4}~yr^{-1}$ in the power spectrum of Titan. The cumulative effects of the Great Inequality and of Titan and Iapetus lead to the large values of the $\eta$ parameters in fig. \ref{etas-sat} and \ref{retro-eta}.
The chaotic evolution of the two satellites does not lead to destabilisation in the timespan of our integration ($10^{8}$ years), however longer simulations are needed to test the long term stability. It is possible that the irregular behaviour takes the two satellites into other chaotic regions ultimately leading to their expulsion from the system. It is also possible that, during their chaotic wandering, they cross the paths of more massive satellites and be collisionally removed. This could have been the fate of possible other satellites which originally populated the region encompassed between $20.94 \times 10^{6}$ km and $22.44 \times 10^{6}$ km.

\section{Evaluation of the collisional evolution}\label{collisional-evolution}

\begin{figure}
\centering
\includegraphics[width=8.0cm]{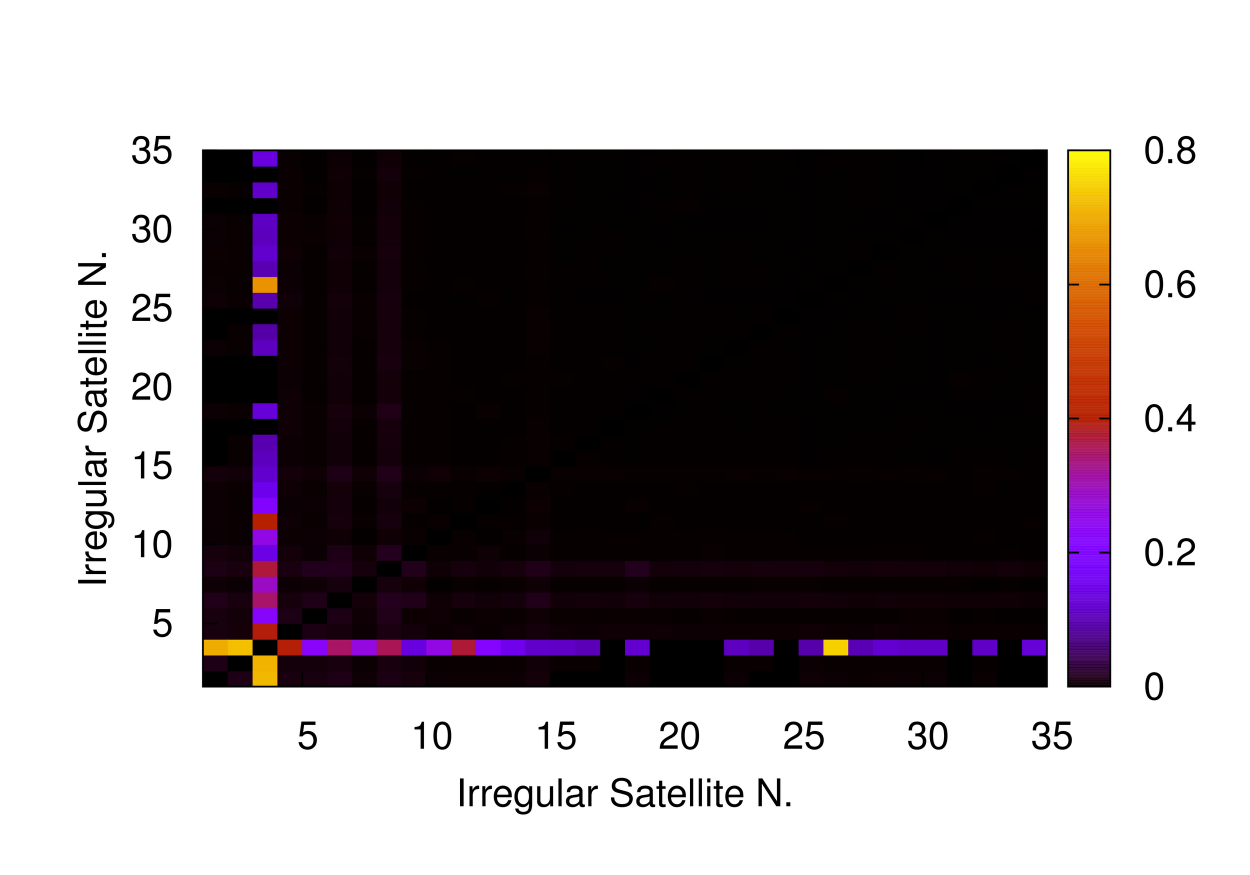}
\caption{Colour map of the collisional probabilities between the $35$ presently known irregular satellites of Saturn. For each pair of satellites the probability of collision is evaluated in terms of the number of impacts during the system lifetime ($3.5 \times 10^9$ years). The colour coding used to illustrate the impact probabilities ranges from yellow (highest values) to blue (lowest values). Black is used when the orbits do not intersect. The only pairs of satellites for which the impact probability is not negligible always include Phoebe as it can be argued by the fact that only column/line $3$ (Phoebe) is characterised by bright colours.}
\label{collisions}
\end{figure}

\begin{figure*}
\centering
\includegraphics[width=16.0cm]{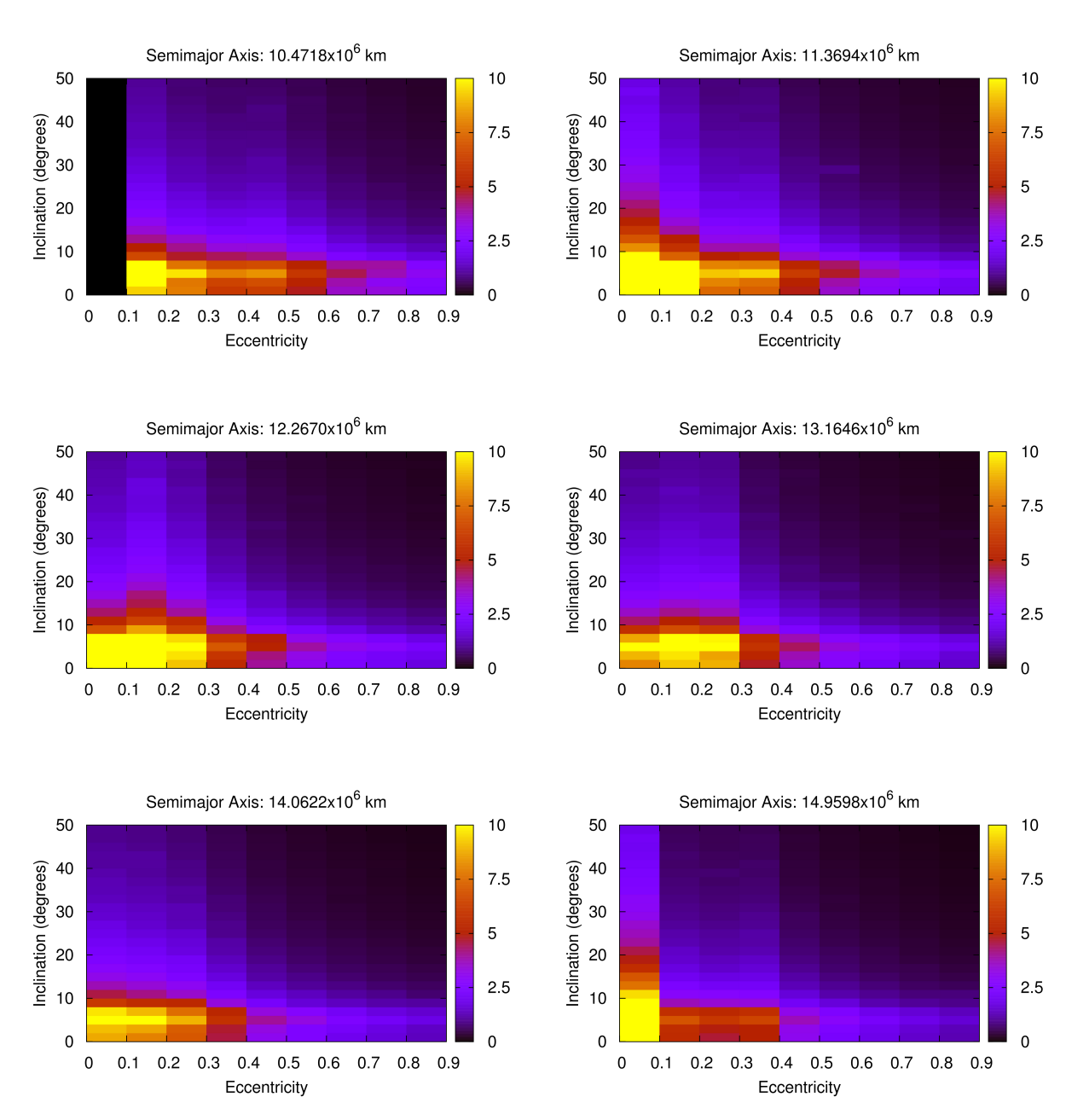}
\caption{Colour map of the collisional probabilities between Phoebe and a cloud of prograde, $3$ km radius test particles filling the region of Phoebe's gap at six values of semimajor axis: $10.47 \times 10^{6},\,11.37 \times 10^{6},\,12.27 \times 10^{6},\,13.16 \times 10^{6},\,14.06 \times 10^{6},\,14.96 \times 10^{6}$ km. The probabilities of collision are evaluated in terms of the number of impacts during the system lifetime ($3.5 \times 10^9$ years) and are represented through a colour code going from yellow (highest values) to blue (lowest values) with black representing the case in which the orbits do not intersect. Note that, due to the high values of collisional probabilities reached in the prograde cases (up to $30$ collisions in the timespan considered) we had to limit the colour range to a maximum of $10$ collisions to maintain the scale readability.}
\label{pro-cloud-3}
\end{figure*}

\begin{figure*}
\centering
\includegraphics[width=16.0cm]{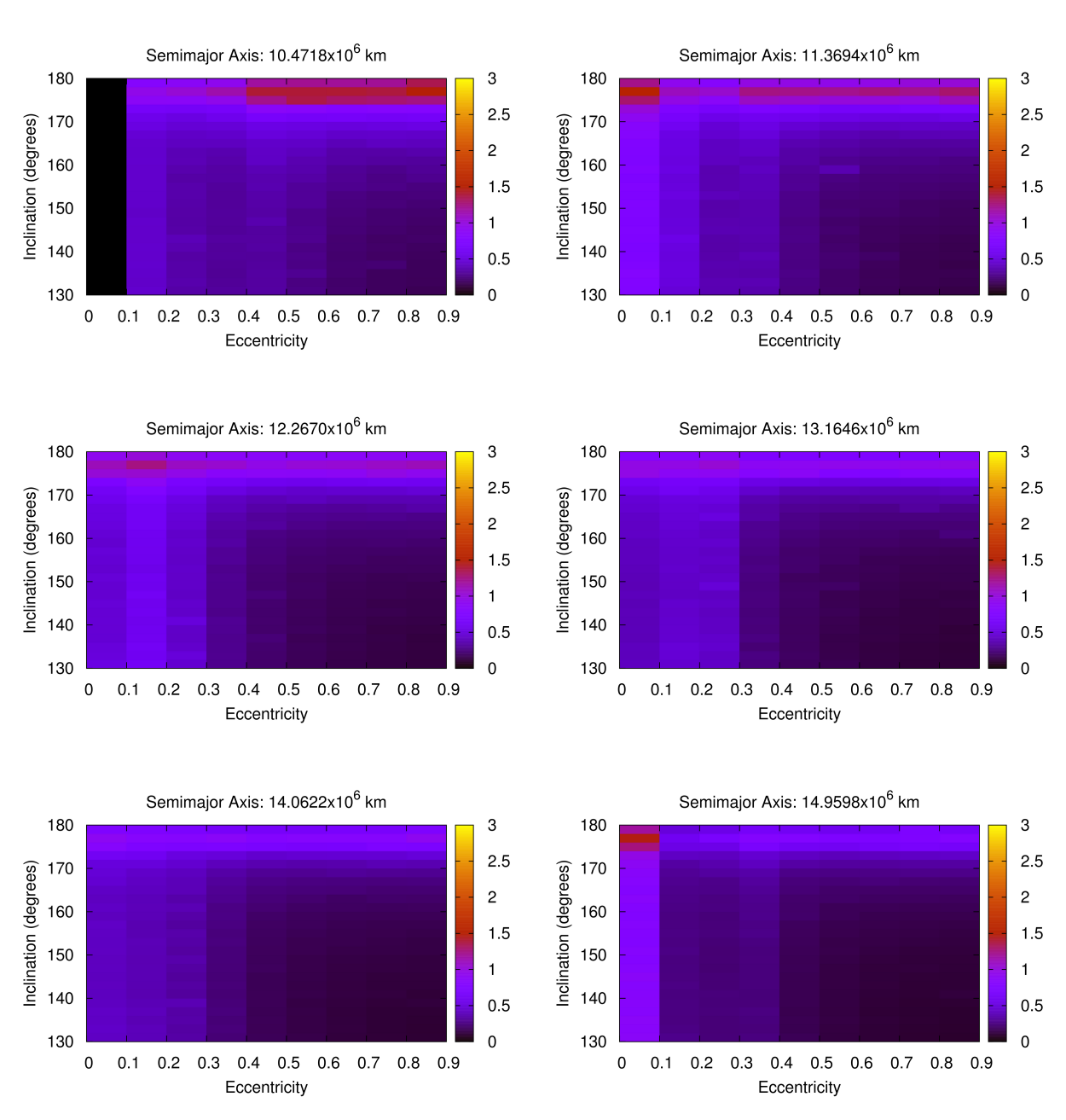}
\caption{Colour map of the collisional probabilities between Phoebe and a cloud of retrograde, $3$ km radius test particles filling the region of Phoebe's gap at six values of semimajor axis: $10.47 \times 10^{6},\,11.37 \times 10^{6},\,12.27 \times 10^{6},\,13.16 \times 10^{6},\,14.06 \times 10^{6},\,14.96 \times 10^{6}$ km. The probabilities of collision are evaluated in terms of the number of impacts during the system lifetime ($3.5 \times 10^9$ years) and are represented through a colour code going from yellow (highest values) to blue (lowest values) with black representing the case in which the orbits do not intersect.}
\label{retro-cloud-3}
\end{figure*}

\begin{figure*}
\includegraphics[width=16.0cm]{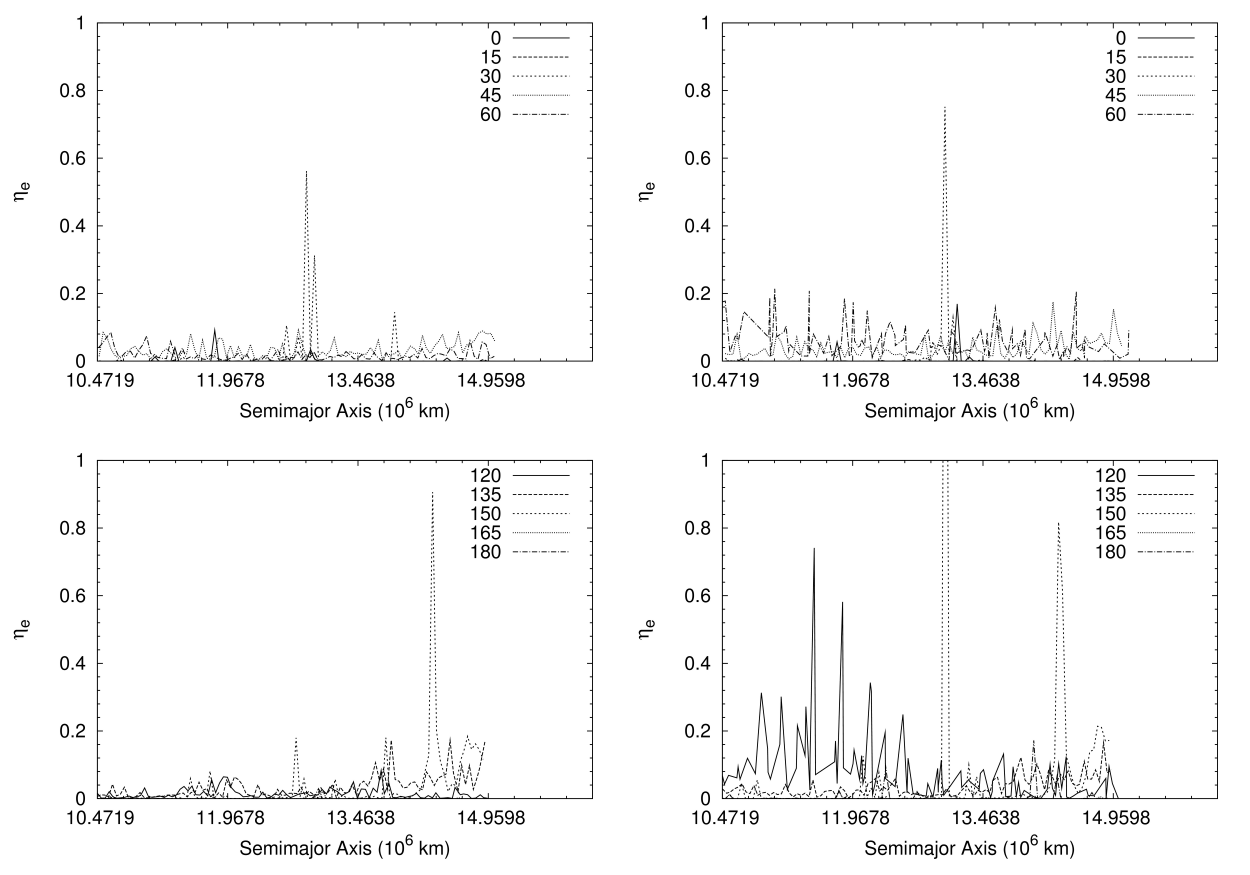}
\caption{Comparison between the $\eta_e$ parameters of prograde (top panels) and retrograde (bottom panels) test particles moving close to the \textit{Phoebe's gap} (i.e., the region extending from $10.47 \times 10^{6}$ km to $14.96 \times 10^{6}$ km from Saturn) in Models $1$ (left column) and $2$ (right column). By comparing the plots pairwise, we note that the differences increase with larger inclinations of the test particles. Those orbits with inclination in the range $30^\circ-60^\circ$ for the prograde case and $120^\circ-150^\circ$ for the retrograde one are the most influenced by the presence of Titan and Iapetus. This hold particularly true in the inner part of this region (i.e. up to Phoebe's orbital distance). Prograde test particles are less affected by the presence of Titan and Iapetus than retrograde ones. In Model $1$, the retrograde test particles have smaller values of $\eta_e$ than the prograde ones. In Model $2$, the retrograde test particles show a marked increase in their $\eta_e$ values for high inclination orbits (i.e. $\sim120^{\circ}$). Such orbits should likely be subject to a strongly chaotic evolution and could be short--lived on a timescale longer than the one ($10^{6}$ years) we considered. A noteworthy feature present in all plots is the peak at $\sim13.015 \times 10^{6}$ km (the orbital distance of Phoebe) for both prograde and retrograde test particles. A second peak, peculiar to retrograde orbits, is located at $\sim14.21 \times 10^{6}$ km.
These peaks appear only for inclination of $30^\circ$ ($150^{\circ}$ for retrograde orbits). These values bring the orbits of the test particles nearer to the equatorial plane of Saturn (once corrected for the giant planet's axial tilt) and thus nearer to the orbital plane of Titan and, to a minor extent, Iapetus. The peak at $\sim13.015 \times 10^{6}$ km is $\sim3$ times lower for retrograde particles than for prograde ones in Model $1$. In Model $2$ the behaviour is opposite and the $\eta_e$ value is increased by the same factor of $\sim3$. Phoebe's evolution has been protected from the effects of these perturbations by its inclination value, which put the satellite into a dynamically safe region.}
\label{phoebe-gap}
\end{figure*}

To understand the present orbital structure of Saturn's satellite system and how it evolved from the primordial one we have to investigate the collisional evolution within the system. Impacts between satellites, in fact, may have removed smaller bodies and fragmented the larger ones. Minor bodies in heliocentric orbits like comets and Centaurs may have also contributed to the system shaping by colliding with the satellites as addressed by \cite{nes04}. At present, however, such events are not frequent because of the reduced flux of minor bodies and the small sizes of irregular satellites \citep{zah03,nes04}.\\
The last detailed evaluation of the collisional probabilities between the irregular satellites around the giant planets was the one performed by \cite{nes03}, which showed that the probabilities of collisions between pairs of satellites were rather low and practically unimportant. The computed average collisional lifetimes were tens to hundreds of times longer than Solar System's lifetime. The only notable exceptions were those pairs involving one of the big irregular satellites (e.g. Himalia, Phoebe, etc.) in the systems. In the Saturn system, Phoebe is between one and two order of magnitudes more active than any other satellite. The authors computed a cumulative number of collisions between $6$ and $7$ in $4.5 \times 10^{9}$ years \citep{nes03}.\\
However, at the time of the publication of the work by \cite{nes03} only $13$ of the $35$ currently known irregular satellites of Saturn had been discovered. We extend here their analysis taking advantage of the larger number of known bodies and of the improved orbital data. Using the mean orbital elements computed with Model $2$ we have estimated the collisional probabilities using the approach described by \cite{kes81}. Since in the scenario described by the Nice Model the Late Heavy Bombardment (LHB) represents a lower limit for the capture epoch of the irregular satellites \citep{gom05,tsi05}, we considered a time interval for the collisional evolution of $3.5 \times 10^{9}$ years, making the conservative assumption that the LHB took place after about $10^{9}$ years since the beginning of Solar System formation. Since the collisional probability depends linearly on time, our results can be immediately extended to longer timescales.\\
The results of our computations (see fig. \ref{collisions}) confirm that the only pairs of satellites with high probability of collisions involve Phoebe. The satellite pairs with the higher number of collisions over the considered timespan are Phoebe--Kiviuq ($0.7126$), Phoebe--Ijiraq ($0.7099$) and Phoebe--Thrymr ($0.6436$). The remaining pairs involving Phoebe have a number of collisions between $0.1$ and $0.35$ (see fig. \ref{collisions}, line/column $3$), with the highest values associated with the prograde satellites Paaliaq, Siarnaq, Tarvos, Albiorix, Erriapo and Bebhionn (S/2004 S11). All the other satellite pairs, due also to their small radii, have negligible ($< 10^{-2}$) collisional probabilities. The predicted total number of collisions in Saturn's irregular satellite system, obtained by summing over all the possible pairs, is of about $12$ collisions over the considered timespan. Half of these impacts involve Phoebe. This is probably at the origin of the gap centred at Phoebe and radially extending from about $11.22 \times 10^{6}$ km to about $14.96 \times 10^{6}$ km from Saturn (i.e. between Ijiraq's and Paaliaq's mean orbits) for both prograde and retrograde satellites. To further confirm this hypothesis we evaluated the impact probability for a cloud of test particles populating this region.\\
We filled with test bodies a volume in the phase space defined in the following way:
\begin{itemize}
\item $10.47 \times 10^{6}\,km \leq a \leq 14.96 \times 10^{6}\,km$
\item $0 \leq e \leq 0.9$
\item $0^{\circ} < i \leq 50^{\circ}$ for prograde orbits 
\item $130^{\circ} \leq i < 180^{\circ}$ for retrograde orbits
\end{itemize}
The sampling stepsize were ${\delta}a=4.488 \times 10^{4}$ km, ${\delta}e=0.1$ and ${\delta}i=2^{\circ}$, for a total of $2860$ prograde orbits and $2860$ retrograde ones. A radius of $3$ km has been adopted for the test particles to compute the cross sections. The results are presented in the colour maps of fig. \ref{pro-cloud-3} for the prograde cases and \ref{retro-cloud-3} for the retrograde ones.
Our results show that lowest collision probabilities (of the order of $0.1$ collisions) are related to orbits with high values of eccentricity and inclination ($e > 0.5$ and $i > 25^{\circ}$ or $i < 155^{\circ}$) as illustrated in the top right quadrants of each plot of fig. \ref{pro-cloud-3} and the bottom right quadrants of each plot of fig. \ref{retro-cloud-3}. Prograde low inclination orbits with $0^{\circ} < i \leq 10^{\circ}$ and $e < 0.5$ have more than $4$ collisions over the given timespan (see the bottom left quadrants of each plot of fig. \ref{pro-cloud-3}). Retrograde orbits with $174^{\circ} < i < 180^{\circ}$ have at least one collision for any value of eccentricity (see the top part of each plot of fig. \ref{retro-cloud-3}). The number of collisions with Phoebe is generally higher for the prograde test particles ($1-5$ collisions) compared to the retrograde orbits ($0.5-1$). These results support the existence of a \textit{Phoebe's gap} caused by collisional erosion. It cannot be due to dynamical clearing mechanisms since, according to an additional set of simulations, we show that the region around the \textit{Phoebe's gap} is stable. $100$ test particles are uniformly distributed
across the \textit{Phoebe's gap} with the following criteria:
\begin{itemize}
\item semimajor axis from $10.47 \times 10^{6}$ km to $14.96 \times 10^{6}$ km
\item radial spacing between the particles of $4.488 \times 10^{4}$ km
\item initially circular orbits
\item initial inclinations set to fixed values equal to $0^{\circ},\,15^{\circ},\,30^{\circ},\,45^{\circ},\,60^{\circ}$ for prograde orbits 
and $120^{\circ},\,135^{\circ},\,150^{\circ},\,165^{\circ},\,180^{\circ}$ for retrograde orbits.
\end{itemize}
The orbits have been integrated for $10^{6}$ years and for each of them a value of $\eta$ has been computed (see Fig. \ref{phoebe-gap}). The $\eta$ values indicate that the perturbations of Titan and Iapetus are negligible and that test particles positioned in regions of high collision probability (i.e. for $i \leq 10^{\circ}$ and $i \geq 174^{\circ}$) are not dynamically unstable.\\
The collisional origin of the Phoebe's gap is also confirmed by observational data showing that the irregular satellites moving closer to Phoebe are those located in regions of the phase space where the impact probability with Phoebe is lower. In addition, the images of Phoebe taken by ISS on-board the Cassini spacecraft revealed a strongly cratered surface, with a continuous crater size distribution ranging from about $50$ m, a lower value imposed by the resolution of ISS images, to an upper limit of about $100$ km comparable to the dimension of the satellite.
A highly cratered surface was predicted by \cite{nes03}, who also suggested that the vast majority of Phoebe's craters should be due to either impacts with other irregular satellites (mainly prograde ones) or be the result of a past intense flux of bodies crossing Saturn's orbit. Comets give a negligible contribution having a frequency of collision with Phoebe of about $1$ impact every $10^{9}$ years \citep{zah03}.
Our results confirm those of \cite{nes03} showing that indeed Phoebe had a major role in shaping the structure of Saturn's irregular satellites. The existence of a primordial now extinct population of small irregular satellites or collisional shards between $11.22 \times 10^{6}$ km and $14.96 \times 10^{6}$ km from Saturn could explain in a natural way the abundance of craters on Phoebe's surface.\\
Phoebe's sweeping effect appears to have another major consequence, related to the existence of Phoebe's gap: it argues against the hypothesis of a Phoebe family. The existence of Phoebe's family has been a controversial subject since its proposition by \cite{gla01}. Its existence was guessed on the close values of inclination of Phoebe and other retrograde satellites. The dynamical inconsistency of this criterion has been pointed out by \cite{nes03}, who showed that the velocity dispersion required to relate the retrograde satellites to Phoebe would be too high to be accounted as realistic in the context of the actual knowledge of fragmentation and disruption processes. Our results suggest also that if a breakup event involved Phoebe, the fragments would have been ejected within the Phoebe's gap for realistic ejection velocities. As a consequence, they would have been removed by impacting on Phoebe.
\FloatBarrier

\section{Existence of collisional families}\label{collisional-families}

\begin{figure}
\centering
\includegraphics[width=8cm]{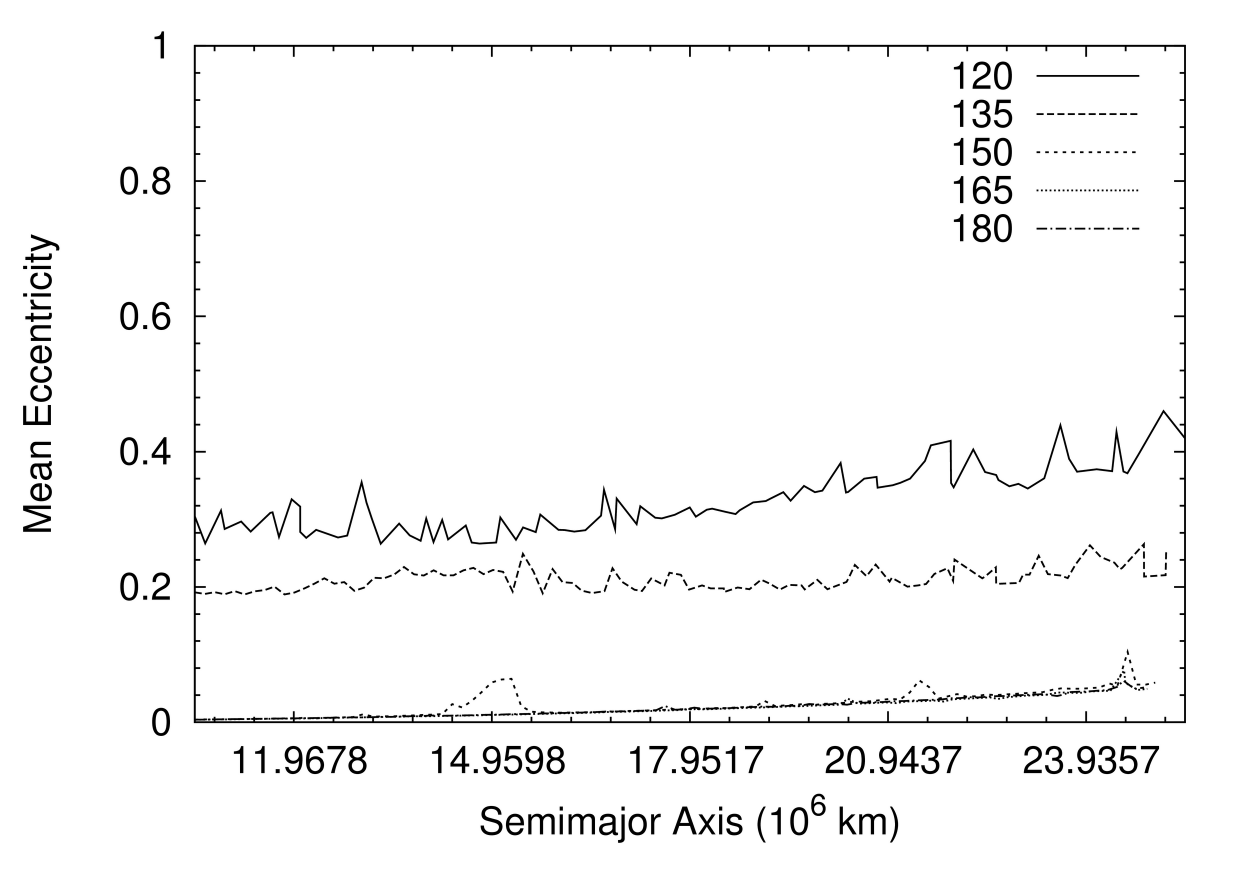}
\caption{Mean eccentricities of the retrograde test particles computed with Model $2$ for different values of inclination. The test particles were on initially circular orbits. Most of the irregular satellites on retrograde orbits lie in the range of inclination where the increase in eccentricity is limited on a $10^{6}$ years timescale. Narvi family is a different case. Narvi spends a significant fraction of its dynamical evolution in the excited region while S/2004 S18 is inside it most of the time. As it can be argued from the plot, the subsequent divergent evolutions could be at the origin of the high velocity dispersion within the group, assuming that the two satellites originated from a common parent body.}
\label{forced-e}
\end{figure}

\begin{table*}
\centering
\begin{minipage}{107mm}
\caption{Dynamical clustering of Saturn's irregular satellites. The term \textit{cluster} used in the table refers to the merging of previously reported families.}
\label{families}
\begin{tabular}{ccc}
\hline
\\
\textbf{Family name} & \textbf{Family members} & \textbf{Dispersion} \\
\\
\hline
\\
 & {\bf Prograde Satellites} & \\ 
\\
Kiviuq   & Kiviuq, Ijiraq & $102$ m/s\\
Albiorix & Albiorix, Erriapo, Tarvos, S/2004 S11 & $129$ m/s\\
Siarnaq & Paaliaq, Siarnaq & $314$ m/s\\
Siarnaq + Albiorix &  Albiorix \& Siarnaq families & $443$ m/s\\
Prograde & All prograde satellites & $532$ m/s\\
\\
& {\bf Retrograde Satellites} & \\
\\
S/2004 S15 & S/2004 S15, S/2006 S1 & 114 m/s \\
Mundilfari & Mundilfari,  S/2004 S13, S/2004 S17 & 116 m/s \\
S/2006 S2 & S/2006 S2, S/2006 S3 & 132 m/s \\
S/2004 S10 & S/2004 S10, S/2004 S12, S/2004 S14 & 144 m/s \\
S/2004 S8 & S/2004 S8, S/2006 S5, S/2004 S16 & 168 m/s \\
Narvi & Narvi, S/2004 S18 & 200 m/s \\
Ymir & Ymir, S/2006 S2 family, & 259 m/s \\
& S/2004 S7, S/2006 S7 & \\
Cluster A & Mundilfari family, S/2006 S6, & 150 m/s \\
& S/2004 S10 family & \\
Cluster B & Cluster A, S/2004 S15, S/2006 S1 & 202 m/s \\
Cluster C & Cluster B, S/2004 S8 family, & 240 m/s \\
& S/2004 S9, S/2006 S4 & \\
Cluster D & Cluster C, Ymir family & 267 m/s \\
Retrograde - Phoebe & All retrograde satellites except Phoebe & 315 m/s \\
Retrograde & All retrograde satellites& 658 m/s \\
\\
\hline
\end{tabular}
\end{minipage}
\end{table*}
The identification of possible dynamical families between the irregular satellites of the giant planets had been a common task to all the studies performed on the subject. The existence of collisional families could in principle be explained by invoking the effects of impacts between pairs of satellites and between satellites and bodies on heliocentric orbits.
The impact rate between satellite pairs is low even on timescales of the order of the Solar System age, with the only exception of impacts amongst the most massive irregular satellites. The gravitational interactions between the giant planets and the planetesimals in the early Solar System may have pushed some of them in planet--crossing orbits. This process is still active at the present time and Centaurs may cross the Hill's sphere of the planets. However, \cite{zah03} showed that the present flux of bodies, combined with the small size of irregular satellites, is unable to supply an adequate impact rate. If collisions between planetesimals and satellites are responsible for the formation of families, these events should date back sometime between the formation of the giant planets and the Late Heavy Bombardment.\\
The possible existence of dynamical families in the Saturn satellite system has been explored by using different approaches. Photometric comparisons has been exploited by \cite{gra03,gra04,bea05} and were limited to a few bright objects. Dynamical methods have been used by \cite{gra03,nes03,gra04} but on the limited sample (about one third of the presently known population) of irregular satellites available at that time. These methods aimed to identify those satellites which could have originated from a common parent body following one or more breakup events. The identification was based on the evaluation through Gauss equations of the dispersion in orbital element space due to the collisional ejection velocities. In this paper we apply the \textit{Hierarchical Clustering Method} (hereinafter \textit{HCM}) described in \cite{zap90,zap94} to the irregular satellites of Saturn. HCM is a cluster--detection algorithm which looks for groupings within a population of minor bodies with small nearest--neighbour distances in orbital element space. These distances are translated into differences in orbital velocities via Gauss equations and the membership to a cluster or family is defined by giving a limiting velocity difference (cutoff).
\cite{nes03} adopted a cutoff velocity value of $100$ m/s according to hydrocode models \citep{ben99}. Here we prefer to relax this value to $200$ m/s considering the possible range of variability of the mean orbital elements because of dynamical effects. The results we obtained are summarised in table \ref{families} and interpreted as in the following.\\
By inspecting our data, we conclude that, as already argued by \cite{nes03}, the velocity dispersion of prograde and retrograde satellites (about $500$ m/s for progrades and more than $600$ m/s for retrogrades) makes extremely implausible that each of the two groups originated by a single parent body.
In addition, the classification in dynamical groups based on the values of the orbital inclination originally proposed by \cite{gla01} and reported by other authors (see \cite{she06} and references within) is probably misleading. Prograde satellites like Kiviuq, Ijiraq, Siarnaq and Paaliaq do share the same inclination but, as a group, they have a velocity dispersion of over $450$ m/s, hardly deriving from the breakup of a single parent body. The enlarged population of retrograde satellites we have analysed show that the clustering around a single inclination \citep{gla01} and their association to Phoebe \citep{gla01} is not an indication of a common origin. The required velocity dispersion is in fact about $650$ m/s.\\
We found two potential dynamical families between the prograde satellites: the couple Kiviuq--Ijiraq and what we term as \textit{Albiorix family}, composed of Albiorix, Erriapo, Tarvos and S/2004 S11. The analysis of the retrograde satellites is more complex. There are three possible groups each composed of three satellites and two others by two satellites, all characterised by acceptable values of the velocity dispersion (between $100$ and $170$ m/s). A sixth possible group satisfying our acceptance criterion is composed of Narvi and S/2004 S18, but its interpretation is quite tricky. This group shows a high velocity dispersion, at the upper limit of our range, but the dynamical evolution of both satellites is uncertain on a timescale of $10^{9}$ years. The orbits of both bodies have the most extreme values of inclination among all retrograde satellites. 
Our numerical experiments with test particles showed that for such bodies the eccentricity is strongly coupled to the inclination (see fig. \ref{forced-e}). In our simulations, initially circular orbits became highly eccentric in less than $10^{6}$ years. It is possible that Narvi and S/2004 S18 had similar orbits in the past which later diverged due to the inclination--eccentricity link.\\
Some of our candidate families merge at higher values of the velocity dispersion forming bigger groups we called \textit{clusters}. The most relevant one is that termed as cluster A in table \ref{families}. It is made of two three-body families and an individual satellite and it is defined at a velocity cut--off of $150$ m/s. At a velocity cut--off of $202$ m/s cluster A merges with the two--body family related to S/2004 S15 forming cluster B. Confirming these dynamical groups by comparing their colour indices is a difficult task because of the limited amount of data available in the literature. The only spectrophotometric data concerning Saturn's retrograde satellites are those of Phoebe and Ymir, which, according to our analysis, are separated by a velocity dispersion of more than $600$ m/s. Phoebe appears to have colours not compatible with any other irregular satellite of the system, supporting our claim that Phoebe is not related to the rest of Saturn's present population of irregular satellites.\\
The situation looks more favourable for prograde families: the colours of three members of the possible Albiorix family are available, with two sets of data for Albiorix itself. The colour data are reported in table \ref{family-colour} with the corresponding $1\sigma$ errors. These data seem to be compatible with the hypothesis of a common origin of the group at a $3\sigma$ level.

\begin{table}
\centering
\begin{minipage}{72mm}
\centering
\caption{Colour indices of three candidate members of the Albiorix family: for each colour index the $1\sigma$ error range is reported \citep{gra03}.}
\label{family-colour}
\begin{tabular}{cccc}
\hline
\\
\textbf{Satellite} & $B-V$ & $V-R$ & $V-I$ \\
\\
\hline
\\
Tarvos  & $0.77 \pm 0.12$ & $0.57 \pm 0.09$ & $0.88 \pm 0.11$ \\
Albiorix& $0.89 \pm 0.07$ & $0.50 \pm 0.05$ & $0.91 \pm 0.05$ \\
        & $0.98 \pm 0.07$ & $0.47 \pm 0.04$ & $0.92 \pm 0.04$ \\
Erriapo & $0.83 \pm 0.09$ & $0.49 \pm 0.06$ & $0.61 \pm 0.12$ \\
\\
\hline
\end{tabular}
\end{minipage}
\end{table}

\section{Conclusions}\label{conclusion}

The aim of this work was to investigate the dynamical and collisional nature of the Saturn system of irregular satellites. We analysed the secular dynamical evolution of the satellites on a timespan of $10^{8}$ years, computing their mean orbital elements. We found evidences of resonant and chaotic behaviours in the motion of about two third of the satellites. We also explored the dynamical features of the phase space close to them and verified the influence of Titan and Iapetus as well as of the Great Inequality in shaping the satellite system.\\
In this paper we have also verified that the present orbital structure is long--lived against collisions but we found indications that in the past a more intense collisional activity could have taken place, mainly due to Phoebe's sweeping effect. By considering the impact rates of the present population, we deduce that the original population could have been at least $\sim 30\%$ more abundant. We also suggest that the absence of prograde and low inclination ($\sim 170^{\circ} - 180^{\circ}$) retrograde irregular satellites in the region encompassed between $10.47 \times 10^{6}$ km and $14.96 \times 10^{6}$ km is a by-product of the sweeping effect of Phoebe. It is less clear if the absence of retrograde satellites with lower inclinations (i.e. high velocities ejecta from Phoebe or other captured bodies) in the same radial region could be due to the same reason or if it is a primordial structure related to the capture mechanism.\\
We also found evidences of dynamical groupings among prograde and retrograde satellites, even if their interpretation in terms of families is not straightforward due to the effects of chaotic and resonant evolution. By applying the HCM algorithm and assuming a velocity cut--off of about $200$ m/s, we retrieved two candidate families between the prograde and six between the retrograde satellites. Some of these families merge in bigger clusters at higher but possibly still acceptable velocity cut--offs. This might be an indication that the system suffered intense post--breakup collisional evolution which could have dispersed the original families.\\

\section*{Acknowledgements}

D.T. wishes to thank David Nesvorny for the fruitful discussions and the help in studying the collisional evolution of the irregular satellites in Saturn system. All the authors wish to thank the anonymous referee for the help and the suggestions to improve this paper.

\bsp

\label{lastpage}

\end{document}